\documentclass{aa}
\usepackage{graphicx}
\usepackage{amsmath}
\usepackage{gensymb}
\usepackage{textcomp}
\usepackage{txfonts}
\usepackage{xcolor}
\usepackage{hyperref}
\hypersetup{
    colorlinks=true,
    citecolor=blue,
    linkcolor=blue,
    filecolor=blue,      
    urlcolor=blue}

\begin{document}

   \title{Age and metallicity of low-mass galaxies: From their centres to their stellar halos}
   \titlerunning{Age and metallicity of low-mass galaxies}

   \author{Elisa A. Tau\thanks{\email{elisa.tau@userena.cl}}
          \inst{1},
          Antonela Monachesi\inst{1},
          Facundo A. Gómez\inst{1},
          Robert J. J. Grand\inst{2},
          Rüdiger Pakmor\inst{3}, 
          Freeke van de Voort\inst{4},
          Federico Marinacci\inst{5,6}, 
          \and 
          Rebekka Bieri\inst{7}}

   \institute{Departamento de Astronomía, Universidad de La Serena, Av. Raúl Bitrán 1305, La Serena, Chile
         \and
         Astrophysics Research Institute, Liverpool John Moores University, 146 Brownlow Hill, Liverpool L3 5RF, UK
        \and
            Max-Planck-Institut f\"ur Astrophysik, Karl-Schwarzschild-Str. 1, D-85748, Garching, Germany
        \and
            Cardiff Hub for Astrophysics Research and Technology, School of Physics and Astronomy, Cardiff University, Queen's Buildings, Cardiff CF24 3AA, UK
        \and
        Department of Physics and Astronomy "Augusto Righi", University of
        Bologna, Via P. Gobetti 93/2, I-40129 Bologna, Italy
        \and
        INAF, Astrophysics and Space Science Observatory Bologna, Via P. Gobetti
        93/3, I-40129 Bologna, Italy
        \and
        Institut f\"ur Astrophysik, Universit\"at Z\"urich, Winterthurerstrasse 190, 8057 Z\"urich, Switzerland
        }
    \authorrunning{Tau et al.}

   \date{Received XXXX; accepted YYYY}

 
  \abstract
   {}
   {In this work we aim to analyse the metallicity and the ages of the stellar halos of low-mass galaxies ($M_* \lesssim 10^{10} \, M_\odot$) to better understand their formation history.}
   {We use a sample of $17$ simulated central low-mass galaxies from the Auriga project with a stellar mass range from $\sim 3 \times 10^8 \, M_\odot$ to $\sim 2 \times 10^{10} \, M_\odot$. We analyse the metallicity and the ages of these galaxies and their stellar halos, as well as the relation between these two properties.}
   {We find that all galaxies have negative radial [Fe/H] gradients, ranging from $-0.18$ to $-0.07$ dex/R$_h$, and that the centres of less massive dwarfs ($M_* < 6.30 \times 10^9 M_\odot$) are in general more metal-poor than those of more massive dwarfs. We find no correlation between the metallicity gradients of the galaxies in dex/R$_h$ and their intrinsic properties, such as their stellar mass or their accreted stellar mass. This implies that the metallicity gradients may not be such a straightforward consequence of the evolution of galaxies in the low-mass regime. However, we find a mild correlation between the metallicity gradients of the galaxies computed in dex/kpc and the galaxies' luminosities. We also find that the dispersion in the mass-metallicity relation found in the stellar halos of low-mass galaxies can be explained with the infall time of their most dominant satellite: at a fixed accreted stellar halo mass, dwarf galaxies that accreted this satellite at later times have more metal-rich accreted stellar halos. This is because satellites that were accreted at later times had more time to chemically evolve and enrich their stellar populations before being disrupted. Regarding the ages of the analysed galaxies, we find a prominent U shape in the radial mean age profiles of $\sim 65\%$ of them, which is mainly driven by the in situ stellar material. This presence of a U shape in the age profiles is due to the combination of the cessation of recent star formation at large radial distances and the merger events these galaxies undergo, which redistribute the stellar material to their outer regions. When focusing on the ages of the stellar halos, we find that more massive ones are older than less massive ones. Additionally, we find a strong correlation between the age and the metallicity of the accreted stellar halos of the galaxies in our sample, such that the more metal-rich ones are younger than the more metal-poor.}
   {Our results show a wide variety in ages and metallicities of low-mass galaxies and their stellar halos, reflecting the complex and non-uniform evolutionary pathways these systems can follow. They also emphasise the importance of studying the outskirts of the galaxies as well as the central regions in order to obtain a complete picture of their evolution histories.}


   \maketitle
%
\section{Introduction} \label{sec:intro}

Metallicity gradients have long been studied as a diagnostic tool to understand galaxy formation and evolution \citep[e.g.][]{Kudritzki2012, Kudritzki2014, Tissera2012, Tissera2022, Sextl2024}. In dwarf galaxies, they reveal star formation histories, feedback, and external influences such as mergers \cite[e.g.][]{BenitezLlambay2016} and their environment \citep[e.g.][]{Leaman2013}. 
Observational studies have found a variety of metallicity gradients among Local Group (LG) dwarf galaxies. For example, \citet{Battaglia2006} studied the Fornax dwarf spheroidal (dSph) galaxy using Very Large Telescope (VLT)/Fibre Large Array Multi Element Spectrograph (FLAMES) spectra and images from the Dwarf Abundances and Radial velocities Team (DART) large programme at ESO, and distinguished two main components based on metallicity: a metal-poor component with a mean metallicity of [Fe/H] $\approx -1.7$, which is found throughout the galaxy, and a more metal-rich component with a mean of [Fe/H] $\approx -1$, which is mainly concentrated within $0.7$ degrees ($\sim 1.68$ kpc) from its centre. This spatial concentration of the metal-rich stars generates a metallicity gradient. \cite{Battaglia2011} extended this type of analysis to the Sextans dSph galaxy, confirming it is a metal-poor low-mass system, with an average metallicity of [Fe/H] $= -1.9$, and that its stars exhibit a wide range of metallicities, indicating a complex chemical enrichment history. The authors also note that the central regions of Sextans are more metal-rich compared to the outer regions, again indicating the existence of a metallicity gradient. More recently, \citet{Munoz2023} reported that both the Large Magellanic Cloud (LMC) and the Small Magellanic Cloud (SMC) exhibit negative metallicity gradients, with the SMC showing a steeper decline in metallicity with radius compared to the LMC. In addition, \citet{Taibi2022} analysed 30 LG dwarfs, finding gradients from $\sim -1.3$ to $0$ dex kpc$^{-1}$, mostly decreasing with radius, but no correlation with the stellar mass, luminosity, star formation timescales, morphology, or distance to the Milky Way (MW) or Andromeda (M31). However, their work shows that the strongest metallicity gradients were found in galaxies that likely experienced past merger events based on their observed structural and kinematic characteristics. This was also found in a numerical work by \cite{BenitezLlambay2016}, where the authors studied dwarf galaxies from the Constrained Local UniversE Simulations (CLUES) and identified two necessary conditions for establishing strong metallicity gradients: (i) the occurrence of a merger that disperses the older, typically metal-poor stars and (ii) the subsequent formation of a more centrally concentrated in situ stellar population. 
Given the apparent role of mergers in shaping the metallicity gradients of dwarf galaxies, it becomes particularly relevant to explore whether the accretion time of their progenitors affects the resulting metallicity of the host galaxy.

Stellar age distributions in dwarf galaxies offer an additional powerful diagnostic of their formation and evolutionary histories. By examining age gradients and the spread of stellar ages across different galactic regions, it is possible to trace episodes of star formation and the imprint of past interactions. In MW-mass galaxies, old and red stars tend to populate the innermost regions, while younger and bluer stars populate larger distances along the discs \citep[e.g.][]{Kepner1999, Chiappini2001}. This indicates `inside-out' stellar formation in massive systems, which is facilitated by the presence of discs that hold star formation at larger radii as a result of gas accreted at later times with higher angular momentum \citep[e.g.][]{Larson1976, Monachesi2012, Barnes2014, Morelli2015, Peterken2020}. In contrast, in low-mass galaxies there seems to be `outside-in' star formation because older stars are observed in their outskirts while younger stellar populations are more centrally concentrated. This has been reported in many observational works of both isolated and non-isolated dwarfs, such as \cite{Tully1996, Gallart2008, delPino2015, Weisz2015, Bettinelli2019, MartinezVazquez2021, Zhuang2021, FerreMateu2021, Liao2023}, and \cite{Cohen2024a}, among others.

One explanation for this outside-in trend is that old stars indeed formed in the outer regions and younger stars formed at smaller galactocentric radii. For this scenario to happen, star formation in the outskirts must be suppressed while continuing in central regions. This suppression likely results from gas depletion or reduced gas density in the outer parts. Simulations suggest that environmental effects during infall into a larger halo, such as ram-pressure stripping \citep[e.g.][]{Mayer2006}, can facilitate this process. However, this cannot explain the outside-in star formation found in some isolated dwarfs. Hence, another way of explaining this trend could be using an inside-out formation scenario and a later redistribution or reshuffling of the stellar material. \cite{Riggs2024} used a sample of $73$ simulated dwarf galaxies from the Marvel and DCJL simulations and found that they generally form their stars inside out, similar to more massive galaxies. However, the oldest stars are often redistributed to larger radii due to fluctuations in the gravitational potential well caused by stellar feedback, leading to an apparent outside-in age gradient at $z = 0$.

Interestingly, some observed low-mass galaxies show a reversal behaviour in their age profiles, with a mean age decreasing from the centre to a certain radius, beyond which the mean age increases \citep[e.g.][]{Williams2009, Mostoghiu2018, Pessa2023, Cohen2024b}. This highlights the complexity of star formation and redistribution processes in low-mass systems, and also raises the question of how stellar ages behave, specifically in the stellar halos, which represent the outermost regions of these galaxies.

All of these internal and external processes that shape the radial metallicity and age profiles can also be responsible for redistributing the in situ stellar material into the outskirts of galaxies. Thus, stellar halos of galaxies hold valuable information regarding the formation history of the galaxies they surround. In particular, the metallicities and ages of the stellar populations in these substructures offer key insights into the timing and nature of accretion events, as well as the chemical evolution of the progenitor satellites that contributed to the stellar halo \citep[see e.g.][]{Gonzalez-Jara2025}. In the MW-mass regime, stellar halos exhibit a wide diversity in their ages, metallicities and accretion histories. Some halos are found to be predominantly old \citep[e.g. that of the MW,][]{Helmi2020}, while others contain younger stellar populations, reflecting more recent accretion activity \citep[e.g. M31,][]{Brown2008}. Their metallicity distributions also vary, with both metal-poor and relatively metal-rich substructures traced back to the accretion of multiple dwarf galaxies \citep[e.g.][]{Helmi2008, Gilbert2014, Belokurov2018, Harmsen2017, Conroy2019, Monachesi2019}. Nonetheless, the in situ material contained in the stellar halos is not as significant in this stellar mass range as it is in dwarf galaxies.

In low-mass galaxies, the inner regions of stellar halos are mainly comprised of this in situ component \citep[][see also \citealt{Gonzalez-Jara2025} for results with the CIELO simulations]{Tau2025}. However, the formation mechanisms and properties of stellar halos in the low-mass regime remain less well understood. This is due to the fact that, until recently, it was not clear whether low-mass systems had these extended structures. Consequently, the study of stellar halos in dwarf galaxies only began in earnest after their recent observational detection \citep[e.g.][]{Longeard2023, Jensen2024, Tau2024, Wheeler2025}. Exploring the metallicity and age distributions in this mass regime provides a pathway to assess whether similar evolutionary trends hold for dwarf galaxies, and to what extent the hierarchical formation model predicted by the $\Lambda$ cold dark matter \citep[$\Lambda$CDM,][]{White1978, Searle1978, Navarro1997} paradigm applies at lower mass scales.

Taking advantage of numerical models, we aim to analyse the stellar material's age and metallicity distributions in the low-mass regime to shed light on the formation mechanisms and the evolution of these systems. In \citet[][hereafter T25]{Tau2025}\label{T25ref}, we analyse the stellar halos of 17 low-mass simulated galaxies and find that the inner regions of these structures are dominated by in situ material. This dominance extends to all radii when considering dwarfs with $M_* \leq 4.5 \times 10^8 \, M_\odot$, and this in situ component of the stellar halos is mostly formed in the inner regions of the galaxies and subsequently ejected into the outskirts during interactions and merger events. Furthermore, we find that the more massive dwarf galaxies ($M_* \geq 6.3 \times 10^9 \, M_\odot$) accrete stellar material until later, which impacts the formation time of the accreted component of the stellar halos. Here, we follow up on the work of \hyperref[T25ref]{T25} and study the stellar populations of the same set of galaxies, mainly focusing on their metallicity and age profiles, as well as on their stellar halos' age and metallicity trends. We also explore whether the accretion time of satellites influences the present-day gradients observed in stellar halos, and whether the age of the halo itself correlates with its properties.

This paper is structured as follows: we present our simulated sample and set our definition for the stellar halo in Sect. \ref{sec:methodology}. In Sects. \ref{sec:metanalysis} and \ref{sec:age} we present our results regarding our metallicity and age analyses, respectively. We analyse radial metallicity and age profiles, as well as global values of these properties. In Sect. \ref{sec:agemet}, we present a relation found for the age and the metallicity of the accreted stellar halos of the galaxies. In Sect. \ref{sec:mechanisms}, we discuss internal and external mechanisms that affect the age and metallicity distribution found in these galaxies. We discuss our results and compare them with observations in Sect. \ref{sec:discussion}. Finally, in Sect. \ref{sec:conclusions} we summarise our findings and list our conclusions. 

\section{Methodology} \label{sec:methodology}

\subsection{The Auriga project} \label{subsec:auriga}

The Auriga project \citep{Grand2017} is a suite of high-resolution cosmological magneto-hydrodynamical zoom-in simulations run with the AREPO code \citep{Springel2010, Pakmor2014}, adopting a $\Lambda$CDM cosmology consistent with Planck results ($\Omega_m = 0.307$, $\Omega_b = 0.048$, $\Omega_\Lambda = 0.693$, $H_0 = 100 \, \rm{h} \, km \, s^{-1} Mpc^{-1}$ and $h = 0.6777$; \citealt{PlanckCollaboration2014}). The simulated halos were selected from the dark-matter-only EAGLE simulation \citep{Schaye2015} to be relatively isolated and the most massive systems in their environments at $z=0$.

The fiducial suite of the Auriga simulations consists of $30$ galaxies, which have been used in numerous studies to analyse MW-mass galaxies and their properties (e.g. \citealt{Monachesi2019} for stellar halos, \citealt{Gargiulo2019} for bulges, \citealt{Vera-Casanova2022, Vera-Casanova2025} for stellar streams, \citealt{Pinna2024} for discs, \citealt{Fragkoudi2025, Lopez2025} for bars, \citealt{Kunder2025} for the inner-central stellar halo). This initial set was later extended to include a lower mass range, including $26$ new central low-mass galaxies ranging from $5 \times 10^9 \, M_\odot$ to $5 \times 10^{11} \, M_\odot$ for the DM halos and from $1.23 \times 10^5 \, M_\odot$ to $2.08 \times 10^{10} \, M_\odot$ for their stellar mass. These simulated galaxies are composed of particles with a resolution of $5 \times 10^4 \, M_\odot$ in DM mass and $\sim 6 \times 10^3 \, M_\odot$ in baryonic mass. 
We refer the reader to \cite{Grand2024} for details. This new set of simulations is the one we use throughout this work to study low-mass galaxies.

\subsection{Sample selection} \label{subsec:sample}

\begin{table*}[!htb]
    \centering
    \caption{Characteristics of the galaxies of our sample.}
    \label{tab:properties}
    \begin{tabular}{lcccccccr}
    \hline
    Auriga & Stellar mass & R$_h$ & c & Scale Height & Mean age & Median [Fe/H] & M$_{\stackrel{\text{acc}}{\text{halo}}}$/M$^*_{halo}$ & M$^*_{halo}$/M$_*$ \\
    & [M$_\odot$] & [kpc] & [kpc] & [kpc] & [Gyr] & dex & $> \, 4 \, R_h$ & \\
    \hline
         1 (L5) &    $2.08 \times 10^{10}$ & 7.736 & 4.55 & 1.26 & 6.77 (8.18) & -0.13 (-0.65) & 0.24 & 0.06 \\
         2 (L1) &    $1.81 \times 10^{10}$ & 5.394 & 7.16 & 1.65 & 5.07 (6.12) &  -0.13 (-0.46) & 0.61 & 0.10 \\
         3 (L4) &    $1.80 \times 10^{10}$ & 6.793 & 7.46 & 1.66 & 4.88 (7.03) &  -0.22 (-0.73) & 0.39 & 0.05 \\
         4 (L3) &    $1.78 \times 10^{10}$ & 5.554 & 4.51 & 1.19 & 6.61 (7.54) &  -0.08 (-0.54) & 0.34 & 0.07 \\
         5 (L7) &    $1.23 \times 10^{10}$ & 5.032 & 5.75 & 1.47 & 4.11 (6.96) &  -0.19 (-0.65) & 0.45 & 0.08 \\
         6 (L0) &    $8.83 \times 10^{9}$ & 1.180  & 1.94 & 0.66 & 6.36 (7.43) &  0.00 (-0.52) & 0.01 & 0.09 \\
         7 (L2) &    $8.29 \times 10^{9}$ & 7.050  & 6.72 & 1.56 & 5.03 (7.15) &  -0.27 (-0.82) & 0.53 & 0.03 \\
         8 (L6) &    $7.89 \times 10^{9}$ & 4.802  & 3.76 & 1.19 & 6.39 (7.51) &  -0.14 (-0.57) & 0.32 & 0.06 \\
         9 (L11) &   $7.02 \times 10^{9}$ & 4.544 & 4.92 & 1.24 & 5.15 (7.41) &  -0.21 (-0.66) & 0.31 & 0.05 \\
         10 (L9) &   $6.30 \times 10^{9}$ & 3.938 & 5.84 & 1.41 & 5.41 (5.28) &  -0.21 (-0.50) & 0.18 & 0.15 \\
         11 (L10) &   $5.34 \times 10^{9}$ & 3.314 & 3.87 & 1.15 & 5.47 (6.30) &  -0.23 (-0.57) & 0.07 & 0.08 \\
         12 (L8) &    $5.09 \times 10^{9}$ & 5.540 & 6.37 & 1.49 & 4.92 (7.35) &  -0.37 (-0.86) & 0.43 & 0.05 \\
         13 (LL2) &   $3.46 \times 10^{9}$ & 3.063 & 2.43 & 0.86 & 4.93 (4.72) & -0.22 (-0.46) & 0.02 & 0.05 \\
         14 (LL9) &   $1.06 \times 10^{9}$ & 3.500 & 3.56 & 1.07& 4.98 (5.70) & -0.52 (-0.78) & 0.04 & 0.04 \\
         15 (LL8) &   $4.54 \times 10^{8}$ & 1.034 & 2.36 & 0.72 & 5.95 (5.39) & -0.37 (-0.61) & 0.00 & 0.09 \\
         16 (LL6) &   $4.53 \times 10^{8}$ & 1.923 & 3.36 & 0.74 & 4.55 (3.93) &  -0.57 (-0.75) & 0.01 & 0.07 \\
         17 (LL11) &  $3.28 \times 10^{8}$ & 3.161 & 4.82 & 1.03 & 4.27 (4.87) & -0.70 (-0.97) & 0.04 & 0.02 \\
    \hline
    \end{tabular}
    \tablefoot{Galaxies are listed in order of decreasing stellar mass. The columns indicate: (1): label used in this work (original Auriga run number), (2): stellar mass, (3): half-light radius, (4): semi-minor axis (c) computed for the ellipsoid with semi-major axes equal to $4 \, R_h$, (5): scale height, (6): mean age within $4 \, R_h$ (total stellar halo), (7): median [Fe/H] within $4 \, R_h$ (total stellar halo), (8): accreted mass fraction of the stellar halo, (9): total stellar halo mass fraction.}
\end{table*}

In this work we use the same sample of simulated dwarf galaxies analysed in \hyperref[T25ref]{T25}. For completeness, here we present a brief description of these objects (but see \hyperref[T25ref]{T25} for details). We consider a subsample of $17$ central low-mass galaxies out of the $26$ available in the simulation, with DM masses ranging from $3.06 \times 10^{10} \, M_\odot$ to $3.73 \times 10^{11} \, M_\odot$ and stellar masses ranging from $3.28 \times 10^8 \, M_\odot$ to $2.08 \times 10^{10} \, M_\odot$. This selection of galaxies was made based on a cut in the number of stellar particles that they have: all these 17 galaxies have more than $10000$ stellar particles at $z=0$, allowing us to study their outskirts. We present the properties relevant for this work in Table \ref{tab:properties}. Galaxies were re-labelled according to their decreasing stellar masses. As a result, Auriga 1 is the most massive galaxy and Auriga 17 is the least massive one. We note that the label assigned to each galaxy in this work is not the same as the one used in \cite{Grand2024}. The original Auriga run number is listed in brackets in the first column next to our label, and an ‘L’ or ‘LL’\footnote{Since the low-mass galaxies available in the Auriga project are separated in two simulation sets according to their DM halo mass, to refer to their original run numbers and not repeat them we added an L to the Auriga number of the galaxies with DM halos of $\sim 10^{11} \, M_\odot$ and two L for those galaxies with DM halos of $\sim 10^{10} \, M_\odot$.} is added to distinguish between DM halo masses of $10^{11} \, M_\odot$ and $10^{10} \, M_\odot$ , respectively. The analysed galaxies were rotated from their originally random configuration in order to have the Z-axis aligned with the angular momentum vector of the central galaxy for all particles.

The galaxies analysed in this work do not have bars and not all of them present well-defined discs: the circularities of Auriga 10, 15 and 16 indicate that they are spheroidal galaxies. Hence, $\sim 82\%$ of the galaxies of our sample have a disc component. Some galaxies, such as Auriga 2, 5 and 8, show clear stellar streams and shell features in their external regions (see \hyperref[T25ref]{T25} for details). Black holes are seeded at a mass of $10^5$ in halos of masses greater than $5 \times 10^{10} \, M_\odot$, so the more massive dwarfs of our simulated sample of galaxies (Auriga 1 to Auriga 12) have a central black hole.

We are particularly interested in analysing the stellar halos of these selected galaxies. Thus, it is important to define this substructure. We consider the same definition based on a spatial selection criterion as in \hyperref[T25ref]{T25}: stellar particles located outside an oblate region are considered to be part of the stellar halo. This region is defined such that the semi-major and intermediate axes are $a = b = 4 \, R_h$, where $R_h$ is the half-light radius. The surface of this ellipsoid is determined by the equation $(\frac{x}{4R_h})^2 + (\frac{y}{4R_h})^2 + (\frac{z}{c})^2 = 1$, where c is the semi-minor axis, and it was used as the inner limit of the stellar halo. The values of c obtained for the whole sample range from $1.94$ kpc to $7.46$ kpc, and they are listed in Table \ref{tab:properties}. We also list in Table \ref{tab:properties} the computed scale height of these galaxies, whose values range from 0.66 kpc to 1.66 kpc and are always lower than those of c, with a mean difference of $\sim 3.5$ kpc. All stellar particles bounded to the galaxy that are located outside of this ellipsoid and up to a distance of $10 \, R_{h}$ in all directions from its centre are considered as part of the stellar halo.

Another crucial definition for this analysis is that of accreted particles. In this work, we define accreted particles as those stellar particles that originated from gas bounded to a different galaxy than the main one (i.e. a satellite galaxy or an external galaxy), regardless of the location of the satellite galaxy. In contrast, we define in situ particles as those stellar particles that were born from gas that was bound to the main galaxy, regardless of whether the gas was from the host galaxy or was provided by a gas-rich satellite during an interaction.

Throughout this work, the total stellar halo refers to the combined in situ and accreted stellar components, whereas the accreted stellar halo includes only the accreted stellar material of this structure. The accreted (in situ) mass fractions of the stellar halos of these galaxies range from $\sim 0$ ($\sim 1$) to 0.61 (0.39), although the majority has a stellar halo accreted mass fraction lower than 0.5, with only two galaxies where this quantity surpasses 0.5 (Auriga 2 and 7). The total stellar halo mass fractions of the galaxies range from $0.02$ (Auriga 17) to $0.15$ (Auriga 10). All of these values are listed in Table \ref{tab:properties}.

\section{Metallicity distributions} \label{sec:metanalysis}

In this Section we analyse the metallicity distribution of the galaxies, as well as their corresponding stellar halos. In Sect. \ref{subsec:metallicity} we characterise the metallicity profiles of the different dwarf galaxies in our sample, comparing the metallicity gradients of their central regions to those of their outskirts. We then analyse the metallicity of the accreted stellar halos and how they are affected by their most dominant progenitor in Sect. \ref{subsec:metacchalo}.

\subsection{Metallicity profiles and gradients} \label{subsec:metallicity}

\begin{figure}[!ht]
   \centering
   \includegraphics[width=0.96\columnwidth]{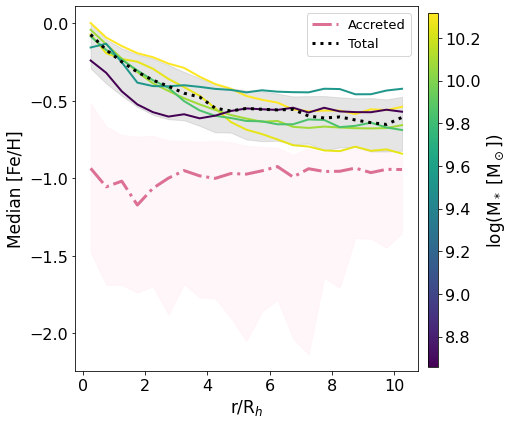}
    \caption{Total median [Fe/H] (dotted black line) and accreted median [Fe/H] (dash-dotted pink line) profiles of the galaxies in our sample. Individual [Fe/H] profiles for six galaxies of our sample are also shown, normalised by their respective $R_h$ and colour-coded by their stellar masses. The shaded regions represent the 16th and 84th percentiles.}
    \label{fig:FeHprof}
\end{figure}

\begin{figure}[!ht]
   \centering
   \includegraphics[width=0.85\columnwidth]{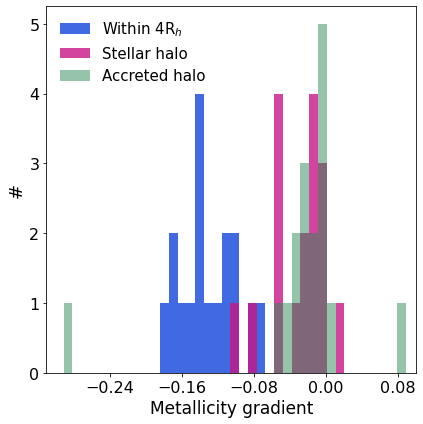}
    \caption{Metallicity gradients in dex/$R_h$ of the galaxies of our sample computed when considering the metallicity profile of the galaxy within $4 \, R_h$ (blue), and the metallicity profile of the total stellar halo (pink) and the accreted stellar halo (green).}
    \label{fig:metgrad}
\end{figure}

\begin{figure*}[!htb]
   \centering
   \includegraphics[width=1.9\columnwidth]{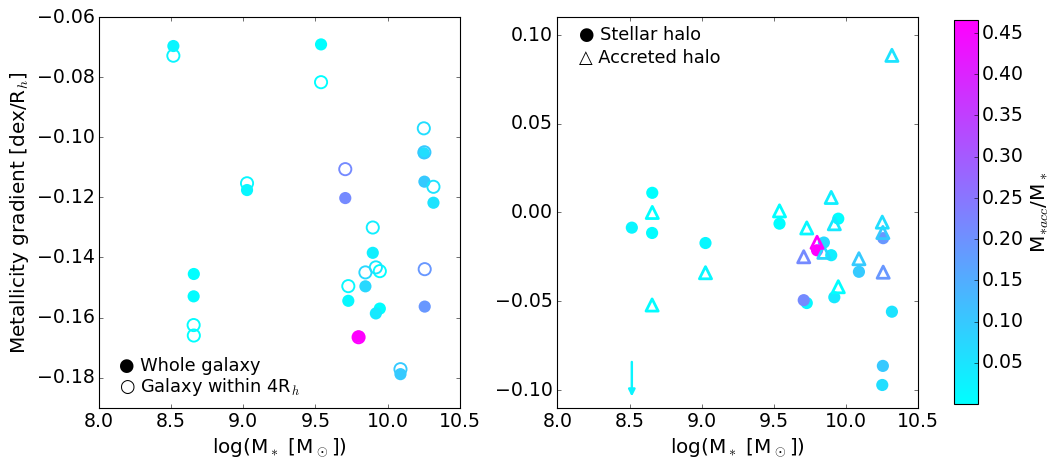}
    \caption{Left panel: Metallicity gradient of the galaxies of our sample computed within $4 \, R_h$ (empty circles) and $10 \, R_h$ (filled circles), as a function of their stellar mass. Right panel: Metallicity gradients of the total stellar halo and the accreted stellar halo as a function of the galaxies' stellar mass, represented with filled circles and empty triangles respectively. The arrow represents the value corresponding to the accreted stellar halo of Auriga 17, which is $\sim -0.3$ dex/R$_h$. Both panels are colour-coded by the galaxies accreted mass fraction.}
    \label{fig:MetGrad_vs_Mstel}
\end{figure*}

In Fig. \ref{fig:FeHprof}, we show with a black dotted line the median of the median [Fe/H] profiles of the galaxies in our sample, computed within concentric 3D ellipsoids along the $R_h$ at $z=0$. The radial distance along the ellipsoid's major axis is normalised by the galaxies' respective $R_h$. We bear in mind that the results obtained in this metallicity analysis are affected by the adopted geometry with which we compute the profiles (see Appendix \ref{appendix0} for the comparison with a cylindrical geometry, mimicking minor axis profiles). The pink dash-dotted line in Fig. \ref{fig:FeHprof} represents the median of the [Fe/H] profiles when considering only the accreted stellar material of the galaxies. We note that this accreted profile is remarkably flat and presents lower values of [Fe/H] at all radii. We also show the individual [Fe/H] profiles for $6$ galaxies of our sample, chosen to span the whole mass range, colour-coded by their total stellar mass. We note that the values of [Fe/H] decrease from the central regions to the outskirts of the galaxies, indicating that the latter are more metal-poor. In addition, we note a trend in this decrease of [Fe/H] such that in the outer regions it is larger for more massive galaxies. This decrease can be quantified by measuring how steeply the values of the metallicity vary within certain regions. In this work, we refer to this variation as metallicity gradient ($\nabla_{\mathrm{[Fe/H]}}$), which is computed as the slope of a weighted linear fit to the median stellar metallicity profile within 10 $R_h$ as a function of normalised radius 

\begin{equation}
\langle \mathrm{[Fe/H]} \rangle (R_i) =
\nabla_{\mathrm{[Fe/H]}} \left( \frac{R_i}{R_h} \right) + b,
\end{equation}

\noindent where $\langle \mathrm{[Fe/H]} \rangle (R_i)$ is the median metallicity measured in the \textit{ith} bin, $R_i$ is the galactocentric radius of that bin, and b is the interceptor. The fit computed with \textit{polyfit} is obtained by minimising a weighted least-squares function,

\begin{equation}
\chi^2 = \sum_i \omega_i \left[ \langle \mathrm{[Fe/H]} \rangle (R_i) - \left( \nabla_{\mathrm{[Fe/H]}} \frac{R_i}{R_h} + b \right) \right]^2,
\end{equation}

\noindent where the weights are defined as $\omega_i = N_i / N_{tot}$, with $N_i$ and $N_{tot}$ being the number of stellar particles in each radial bin and the total number of stellar particles, respectively. The values of the metallicity gradients in dex/$R_h$ of the galaxies in our sample range from $-0.18$ (Auriga 5) to $-0.07$ (Auriga 13). Higher values of metallicity gradients indicate that the metallicity profile is rather flat, while lower values imply that the metallicity profile is steeper. Since the [Fe/H] profiles do not decrease strictly linearly from the centres to the outskirts, we also computed the metallicity gradients for the galaxy (i.e. particles within $4 \, R_h$) and the stellar halo (i.e. between $4$ and $10 \, R_h$) separately. 
The metallicity gradients normalised by the $R_h$ of these galaxies within $4 \, R_h$ range from $-0.18$ (Auriga 5) to $-0.07$ (Auriga 17), while in their stellar halos they range from $-0.1$ (Auriga 4) to $0.01$ (Auriga 15) when considering both the in situ and accreted components, and from $-0.29$ (Auriga 17) to $0.09$ (Auriga 1) when considering only the accreted component. In Fig. \ref{fig:metgrad}, we show the distribution of the gradients within $4 \, R_h$ in blue, and those of the stellar halo and its accreted material in pink and green, respectively. We see that, indeed, the [Fe/H] profile radially decreases more steeply within $4 \, R_h$ than it does in the stellar halo. This reflects the fact that the main bodies of the galaxies undergo significant chemical enrichment, producing strong central–outer metallicity differences within $4 \, R_h$, while the stellar halos are expected to consist mainly of older material, partly associated with old accretion events. 
Notice however that, as shown in Fig. \ref{fig:FeHprof}, the accreted component is typically more metal-poor than the overall stellar halo population. This indicates a significant  contribution of in situ material to the outermost regions of these low mass galaxies.

Given the variety of metallicity gradient values, it is interesting to see if there is any relation between them and the galaxies' properties. We analyse these values as a function of the galaxies' stellar mass, and we show these results in Fig. \ref{fig:MetGrad_vs_Mstel} (see Appendix \ref{app} for these results of the metallicity gradients in physical units, i.e. dex/kpc). Symbols are colour-coded by the galaxies' accreted mass fraction. The left panel of Fig. \ref{fig:MetGrad_vs_Mstel} shows the metallicity gradient of the whole galaxy (i.e. $< 10 \, R_h$, filled circles) and within $4 \, R_h$ (empty circles) considering its in situ and accreted components, while the right panel shows the metallicity gradient of the total stellar halo (circles) and the accreted stellar halo (triangles) separately. We find no correlation between these gradients and the stellar mass or the accreted mass fraction of the galaxies (although a slight correlation is observed when the metallicity gradients are derived in physical units; see Appendix \ref{app}). 
We also analysed these metallicity gradient values as a function of the total amount of accreted mass and the number of significant progenitors (not shown here), but again we found no correlation between these quantities. Thus, the metallicity gradients of these galaxies do not seem to be a direct consequence of their evolution pathways. This is opposite to what is found in MW-like models, where a strong correlation is obtained between the metallicity gradient of the accreted component and the number of significant progenitors \citep{Monachesi2019}. We further explore this distribution in the following section.

\subsection{The metallicity of the accreted stellar halo} \label{subsec:metacchalo}
 
\begin{figure}[!ht]
   \centering
   \includegraphics[width=0.9\columnwidth]{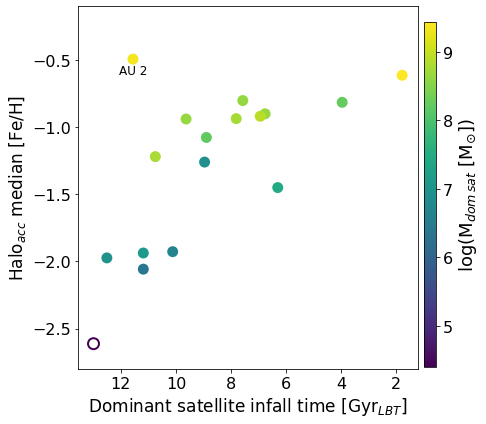}
    \caption{Median [Fe/H] of the accreted stellar halo of the galaxies as a function of their most dominant satellite's infall time, colour-coded by these satellites' stellar mass. The open circle represents Auriga 15, a galaxy that has only a few accreted particles and must therefore be considered with caution.}
    \label{fig:HaloFeH_DomSatInf}
\end{figure}

We now explore the correlation between the most dominant contributor's infall time and the median metallicity of the stellar halo. This allows us to explore whether the infall time of these satellites influences their metallicities. The infall time accounts for the time at which the satellites crossed the $R_{200}$ of the main galaxy. 
In Fig. \ref{fig:HaloFeH_DomSatInf}, we show the median [Fe/H] of the accreted stellar halos of the galaxies as a function of the infall time of their most dominant satellite in terms of stellar mass, which is the destroyed accreted satellite that contributed the most to the stellar halo's accreted component. We note a correlation between these two quantities such that the more metal-rich accreted stellar halos accreted their most dominant satellite at later times. Symbols are colour-coded by the stellar mass of these most dominant satellites, and we can see that the more metal-rich accreted stellar halos have a more massive dominant satellite, as expected. We note that Auriga 2 is an exception to this trend given that it underwent a major merger at early times and accreted a very massive most dominant satellite contributing with a high metallicity. The open circle in Fig. \ref{fig:HaloFeH_DomSatInf} represents Auriga 15, a galaxy with an accreted stellar mass of $M_{acc} = 9.74 \times 10^4 \, M_\odot$, which means that it has only a few accreted particles (for further discussion, see \hyperref[T25ref]{T25}).

\begin{figure*}[!ht]
    \sidecaption
   \includegraphics[width=12cm]{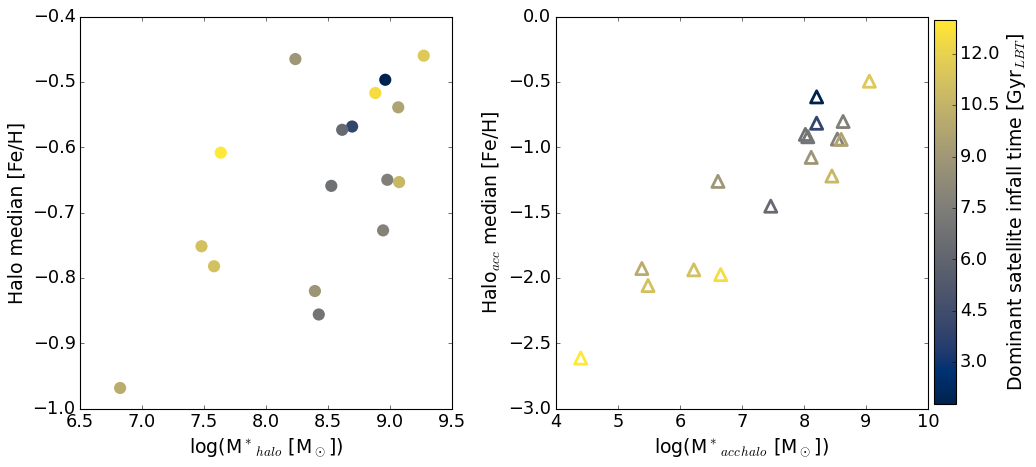}
    \caption{\textit{Left panel}: Median [Fe/H] of the stellar halos of the galaxies as a function of their masses. \textit{Right panel}: Median [Fe/H] of the accreted stellar halos of the galaxies as a function of their masses. Both panels are colour-coded by the mean infall look-back time of the significant progenitors of the stellar halos. We note a correlation between the metallicity of the accreted stellar halo and the infall time of its significant progenitors: more metal-rich accreted halos have accreted their significant progenitors at later times.}
    \label{fig:HaloFeH}
\end{figure*}

We now examine how the median metallicity of these stellar halos connects with their total and accreted stellar halo masses, while also considering the infall time of their most dominant satellite. 
In Fig. \ref{fig:HaloFeH} we show the median [Fe/H] computed for the total stellar halo (left panel) and the accreted stellar halo (right panel) as a function of the total stellar halo's mass and the accreted stellar halo's mass, respectively. Both panels are colour-coded by the 
infall time of the most dominant satellite. We note a wider range of median [Fe/H] of the accreted halo (right) with respect to the overall population (left),  evidencing the significant contribution of in situ material. As already presented in \hyperref[T25ref]{T25}, there is a tight correlation, although with some scatter, between the median [Fe/H] values and the stellar halos' masses when considering the accreted component: more massive accreted stellar halos are also more metal rich. As discussed in \hyperref[T25ref]{T25}, this is due to the fact that more massive stellar halos mainly form through the accretion of more massive progenitor galaxies which, due to the mass-metallicity relation for dwarf galaxies \citep{Kirby2013}, were more metal-rich at the time they were accreted. Interestingly, we also find that the scatter in this relation is due to the different infall times of their most dominant satellites: at a given accreted stellar halo mass, the galaxies that accreted their dominant satellites at later times have a more metal rich accreted stellar halo. This can be explained by the fact that satellites accreted at later times had more time to chemically evolve and enrich their stellar populations prior to disruption, which will then contribute to the accreted stellar halo [Fe/H]. This trend is not found in the left panel, since the mass and metallicity of the total stellar halo are also influenced by the in situ component.

\section{Age analysis} \label{sec:age}

Studying the ages of the galaxies' stellar populations can shed light on their formation history, provide further insight into their chemical enrichment histories, and help determine whether a relation exists between the age and metallicity of the stellar halos. This is an observationally challenging task since deriving stellar ages from the galaxies' stellar populations is difficult. Fortunately, this can be done with simulations, as every stellar particle provides information about their age. 
We characterise the radial age profiles of the galaxies of our sample in Sect. \ref{subsec:ageprof}, specifically focusing on their internal regions. In Sect. \ref{subsec:stelhaloage}, we analyse the mean age of both the total and the accreted stellar halo. 

\subsection{Radial age profiles} \label{subsec:ageprof}

\begin{figure}[!ht]
   \centering
   \includegraphics[width=0.9\columnwidth]{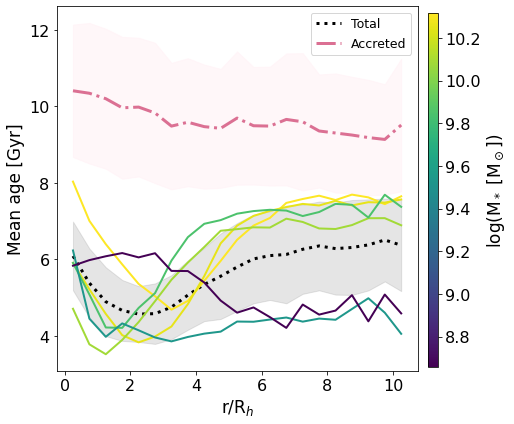}
    \caption{Total mean age profile (dotted black line) and accreted mean age profile (dash-dotted pink line) of all of the galaxies of our sample. The shaded regions represent the standard deviation of these computed means. Individual age profiles for six galaxies of our sample are also shown, normalised by their respective $R_h$ and colour-coded by the galaxies' stellar masses. We note a prominent U shape in the total mean of the age profile of our galaxies, driven by the in situ stellar population.}
    \label{fig:ageprof}
\end{figure}

\begin{figure*}[!ht]
    \sidecaption
   \centering
   \includegraphics[width=12cm]{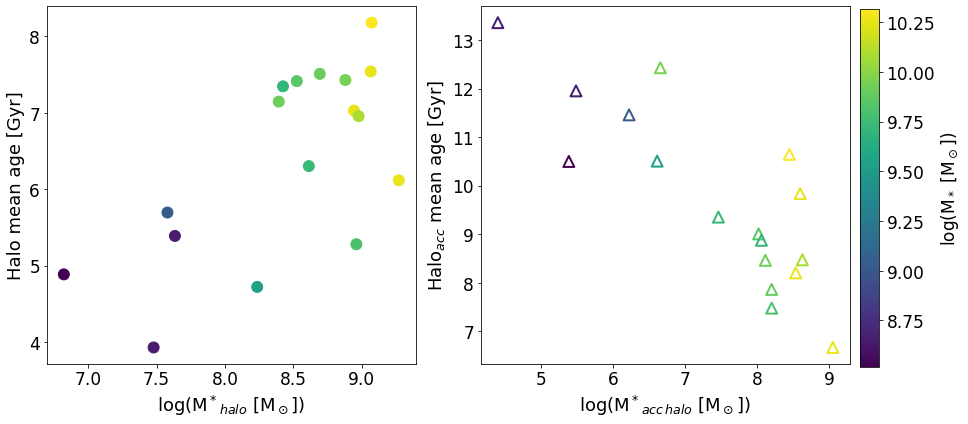}
    \caption{Left panel: Mean age of the total stellar halo as a function of the total stellar halo mass. Right panel: Mean age of the accreted stellar halo as a function of its mass. Both panels are colour-coded by the total stellar mass of the galaxies. We note a correlation in both panels, although they are inverted: more massive stellar halos are older than less massive ones, while more massive accreted stellar halos are younger than the less massive ones.}
    \label{fig:haloage_mhalo}
\end{figure*}

In Fig. \ref{fig:ageprof}, we present the total (in situ + accreted) mean and the accreted mean of the mean radial age profiles of the low-mass galaxies of our sample, shown as black dotted and pink dash-dotted lines, respectively. The shaded regions represent the standard deviation of these computed means. To compute the individual profiles, we used a set of concentric 3D ellipsoids defined at $z=0$, where the semi-major axis of each ellipsoid was scaled by the galaxy’s $R_h$. Within each region, we calculated the mean stellar age by averaging over all stellar particles enclosed in it, or only over the accreted ones for the accreted mean age profile. As expected, we find that the age distribution of the accreted component is significantly older than the overall stellar population distribution. The difference can be as large as 5 Gyr within the inner regions, but becomes smaller ($\sim 3$ Gyr) in the outskirts ($r > 7 \, R_h$). This is because the ages of the in situ populations increase with galactocentric distance, narrowing the difference with the ages of the accreted populations at large distances. We also show the individual total age profiles for $6$ galaxies of our sample (the same ones shown in Fig. \ref{fig:FeHprof}), colour-coded by their stellar masses (see Appendix \ref{appendixA} for the individual profiles of the remaining galaxies). Analysing these individual profiles, we find no correlation between the ages found at different radii and the stellar mass of the galaxies. Nonetheless, the innermost regions of the lower-mass objects in our sample tend to be younger than their more massive counterparts. 
 
Interestingly, we find a prominent U shape in the mean of the total radial age profile of all galaxies (Fig. \ref{fig:ageprof}). This reflects the behaviour of the individual radial age profiles, where 12 of the 17 galaxies in our sample present prominent U-shaped profiles. An inspection of the individual age profiles reveals a general trend: U-shaped profiles tend to appear in the more massive dwarf galaxies, while the less massive ones generally do not exhibit this feature (see Fig. \ref{fig:AgeProfInsAcc}). 
The age profiles for the in situ and accreted stellar particles, computed separately, show that this U-shaped behaviour is associated with the in situ component. This can also be inferred from the lack of this feature in the mean of the accreted age profiles shown in Fig. \ref{fig:ageprof}. 

As seen in \hyperref[T25ref]{T25}, the outskirts of these galaxies are mostly comprised of in situ stellar material. Hence, this upturn in the radial age profiles naturally leads us to examine the ages of the stellar halos themselves.

\subsection{Ages of stellar halos} \label{subsec:stelhaloage}

Building on the results from the radial age profiles, we now turn our focus to the stellar halos as distinct components of the galaxies. To obtain a more complete picture of these galaxies' evolution (particularly in their outer regions) it is also important to explore how the age of the stellar halo relates to its metallicity and its stellar mass. We note in Fig. \ref{fig:ageprof} that the mean of the total age profiles tends to increase from $\sim 4 \, R_h$, indicating that the stellar halo is comprised of old stellar material. This can be quantified by computing the mean stellar age of the stellar halo. We obtained this value for the total stellar halo and for the accreted stellar halo, which are shown in Fig. \ref{fig:haloage_mhalo} as a function of the total stellar halo mass (left panel) and the accreted stellar halo mass (right panel), respectively. Both panels are colour-coded by the galaxies' total stellar mass. We note that the mean ages spanned by the total stellar halo, ranging from $\sim 4$ to $\sim 8$ Gyr, are narrower than those spanned by the accreted stellar halos, which range from $\sim 6$ to $\sim 13$ Gyr. 
Additionally, we find a clear correlation in both cases, although inverted: on one hand, when considering the total stellar halo, we see that more massive stellar halos are older than less massive ones; on the other hand, when considering only the accreted component of the stellar halo, we see that more massive accreted stellar halos are younger than less massive ones.

\begin{figure*}[!ht]
    \sidecaption
   \centering
   \includegraphics[width=12cm]{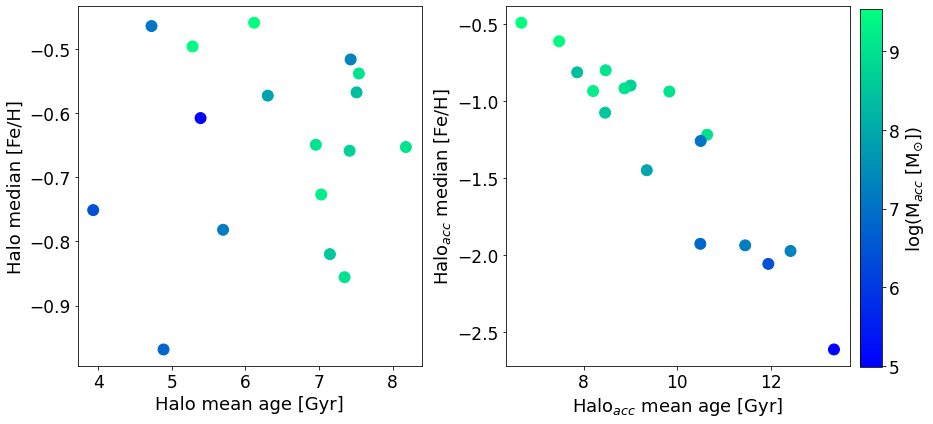}
    \caption{Age-metallicity relation of the galaxies in our sample, colour-coded by their accreted stellar mass. Left panel: Median [Fe/H] of the total stellar halo (in situ $+$ accreted) as a function of its mean age. Right panel: Median [Fe/H] of the accreted stellar halo as a function of its mean age. We note a strong correlation between the median [Fe/H] and the mean age of the accreted stellar halo, such that the more metal-rich accreted stellar halos are also the younger ones.}
    \label{fig:AgeMetRel}
\end{figure*}

The correlation seen in the left panel of Fig. \ref{fig:haloage_mhalo} between the age and mass of the total stellar halo suggests that less massive galaxies keep on forming stellar material until later times, which can then end up comprising their stellar halos. Regarding the correlation in the right panel of Fig. \ref{fig:haloage_mhalo}, we can infer that the less massive galaxies accreted their bulk of stellar material earlier than the more massive ones, leaving them with an older accreted component in their stellar halos. This is a clear consequence of the results of \hyperref[T25ref]{T25}, where it was found that more massive galaxies keep on accreting stellar material until later times, which allows the accreted satellite galaxies to keep on forming stellar material prior to accretion for longer periods so as to then provide the host galaxies with younger stars that end up comprising their accreted stellar halo.

This difference in the accretion time of the satellites also allows them to keep chemically enriching their stellar populations, which has an impact on their metallicity and the resulting metallicity of the stellar halos, as seen in the right panel of Fig. \ref{fig:HaloFeH}. Thus, this could hint at a possible correlation between the age and the metallicity of the stellar halos. We explore this in the following section.

\section{Age-metallicity relation of the stellar halos} \label{sec:agemet}

The chemical enrichment of stellar populations is closely linked to their age: younger stars tend to be more metal-rich because they form from interstellar material that has been progressively enriched by previous generations of stars \citep[e.g.][]{Tinsley1980, Matteucci2003}. Consequently, if more massive galaxies accrete satellites at later times, when these satellites are more chemically evolved, they may end up with younger, more metal-rich accreted stellar halos \citep[e.g.][]{Font2006, Cooper2010}. This naturally suggests a possible correlation between the age and metallicity of stellar halos, which has been studied in MW-mass galaxies \citep[e.g.][]{Das2020, OrtegaMartinez2022}.

The youngest total stellar halo of our sample is that of Auriga 16, with a mean age of $3.93$ Gyr, while the oldest one is that of Auriga 1, with a mean age of $8.18$ Gyr. The mean ages for the total stellar halos of all galaxies are listed in the fourth column of Table \ref{tab:properties}. We find that these structures are, on average, older than the main body of the galaxy (i.e. within $4 \, R_h$), with ages ranging from $4.11$ Gyr to $6.77$ Gyr (see Table \ref{tab:properties}). When analysing the median [Fe/H] of the galaxy's main body and that of the total stellar halos, we find that the main body is more metal-rich than the total stellar halos, with median [Fe/H] values ranging from $-0.70$ dex to $0.00$ dex (see Table \ref{tab:properties}). 
In Fig. \ref{fig:AgeMetRel}, we show the median [Fe/H] of the total stellar halo as a function of its mean age (left panel) and the median [Fe/H] of the accreted stellar halo as a function of its mean age (right panel). Both panels are colour-coded by the galaxies' total accreted stellar mass. We note a clear correlation when it comes to the accreted stellar halos: those that are more metal-rich are also younger, whereas the accreted stellar halos that are more metal-poor are older. This correlation is not present when considering all of the stellar halos' stellar material (i.e. the in situ and accreted components together). This is likely due to the fact that in situ stars dominate the stellar halos of dwarf galaxies in terms of mass (\hyperref[T25ref]{T25}), and their metallicities are primarily governed by the galaxy’s internal chemical enrichment history. As a result, the in situ contribution erases the clear age–metallicity trend observed when considering the accreted halo alone, where the age of the accreted stellar component is affected by the time the satellites stop evolving and are disrupted. This implies that there is a wide diversity in chemical enrichment histories of the in situ populations.

\section{Mechanisms affecting the age and metallicity} \label{sec:mechanisms}

In this section, we explore the mechanisms that affect the redistribution of these dwarf galaxies' stellar material throughout all radii. In Sect. \ref{subsec:sfr}, we address internal processes of the galaxies, such as their star formation rate and stellar migration, while in Sect. \ref{subsec:mergers} we focus on external processes such as merger events.

\subsection{Star formation rate and physical internal processes} \label{subsec:sfr}

Internal processes that are intrinsic to galaxies influence the distribution of their stellar material over time. Such redistribution has a direct impact in the radial [Fe/H] and age profiles we presented in Sects. \ref{subsec:metallicity} and \ref{subsec:ageprof}, respectively. The specific star formation rate (sSFR) is a key diagnostic of how efficiently galaxies are forming stars relative to their stellar mass, and its evolution provides valuable insight into their growth and star formation histories. Variations in the sSFR, both globally and radially, can directly influence the age and metallicity distributions of galaxies. Therefore, studying the sSFR allows us to connect present-day radial profiles with the past star formation activity of galaxies. With this in mind, we investigate the sSFR of our sample to determine whether systematic differences exist between galaxies with and without U-shaped radial age profiles, and how this could affect the metallicity distribution.

One possible explanation for the U-shaped behaviour of the age profiles shown in Fig. \ref{fig:ageprof} is that the star formation of the galaxies presenting this feature is substantially different from that of the galaxies without it. To explore this, we select a subsample of 6 galaxies: the Auriga 1, 3, 5 and 9, all with U-shaped profiles. Conversely, we select Auriga 13 and 15 as cases without U-shaped age profiles. These $6$ galaxies are the same ones for which individual profiles are shown in Figs. \ref{fig:FeHprof} and \ref{fig:ageprof}. Analysing the overall sSFR evolution of our galaxies (not shown here), we find no significant differences between systems with U-shaped radial age profiles and those without them. Galaxies lacking a U-shaped profile generally exhibit smooth star-formation histories with a continuous, approximately monotonic increase in sSFR towards $z=0$. In contrast, some galaxies with U-shaped age profiles show a more bursty behaviour; however, this is not a universal property, as several of them (e.g. Auriga 11 and Auriga 17) display sSFR histories similar to those of galaxies without U-shaped profiles. Likewise, while a subset of U-shaped systems (e.g. Auriga 1, Auriga 3, and Auriga 6) undergo noticeable star-formation bursts, such events are not present in all cases. Therefore, the qualitative differences observed in individual sSFR histories do not translate into systematic population-level differences, and we cannot consider these features as defining characteristics of U-shaped age profiles. 

\begin{figure}[!ht]
   \centering
   \includegraphics[width=0.9\columnwidth]{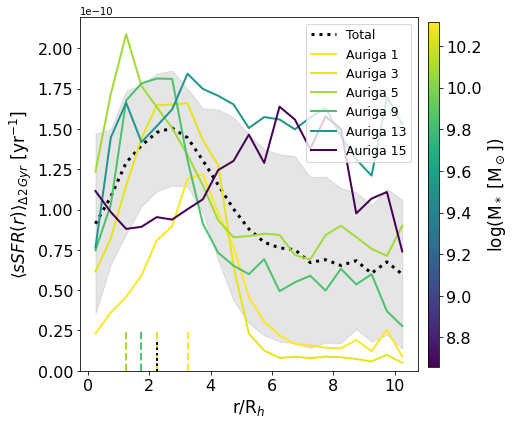}
    \caption{Mean of the $\bigl \langle sSFR(r) \bigr \rangle _{\Delta 2 \, \rm{Gyr}}$ profile (dotted black line) of all the  galaxies in our sample. The individual $\bigl \langle sSFR(r) \bigr \rangle _{\Delta 2 \, \rm{Gyr}}$ profiles the subsample of six galaxies are also shown, colour-coded by their stellar mass. The vertical dashed lines represent the radius at which the minimum age of the U-shaped age profiles in Fig. \ref{fig:ageprof} is located. We note that the radius at which the peak of the $\bigl \langle sSFR(r) \bigr \rangle _{\Delta 2 \, \rm{Gyr}}$ profile occurs matches the radial location of the minimum age of the U-shaped age profiles.}
    \label{fig:sfrprof}
\end{figure}

We now explore radial profiles of the recent, time-averaged star formation rate (SFR). These profiles are computed from the stellar mass formed in situ within the last 2 Gyr in each radial bin, divided by the corresponding time interval and normalised by the total stellar mass enclosed. This provides a proxy for the recent ($\leq 2$ Gyr) star formation activity. 
The black dotted line in of Fig. \ref{fig:sfrprof} represents the mean of all of the $\bigl \langle sSFR(r) \bigr \rangle _{\Delta 2 \, \rm{Gyr}}$ profiles of our sample, the shaded region represents the standard deviation of this mean, and the coloured solid lines represent the individual profiles of our subsample of $6$ galaxies. 

We can clearly see that the peak of the mean of all of the $\bigl \langle sSFR(r) \bigr \rangle _{\Delta 2 \, \rm{Gyr}}$ profiles occurs at the same radial distance at which the minimum of the mean of the age profiles is found, as expected, which is shown by the vertical black dotted line. We also show that the same happens for the 4 individual galaxies we present in Fig. \ref{fig:sfrprof} as examples of galaxies with U-shaped age profiles, for which radii at which their age profiles reach the youngest age is also shown with vertical dashed lines. Thus, we find that the radial age profiles show an upturn beyond the radii at which these galaxies have a sharp drop in their recent star formation. This agrees with previous findings in MW-mass galaxies, such as the work of \cite{VarelaLavin2022} who used a sample of simulated late-type galaxies from the EAGLE simulations and found U-shaped age profiles to be the most common. They show that the radius at which the minimum of the U shape occurs matches the drop of young stellar populations. \cite{Debattista2017} and \cite{Sanchez-Blazquez2009} also explained this behaviour in the age profiles of simulated MW-mass galaxies in the same way (see Sect. \ref{subsec:discussion_age} for further discussion). As we also show in Fig. \ref{fig:sfrprof}, this behaviour is not present in the $\bigl \langle SFR_{\Delta 2 \, \rm{Gyr}}\bigr \rangle$ profiles of those galaxies without a U-shaped age profile. In those cases, there is recent star formation ongoing at large distances. 
However, it remains to be explained why the age of the galaxies of our sample with U-shaped profiles rise at large radii, rather than simply flattening. 

\begin{figure*}[!ht]
   \centering
   \includegraphics[width=1.87\columnwidth]{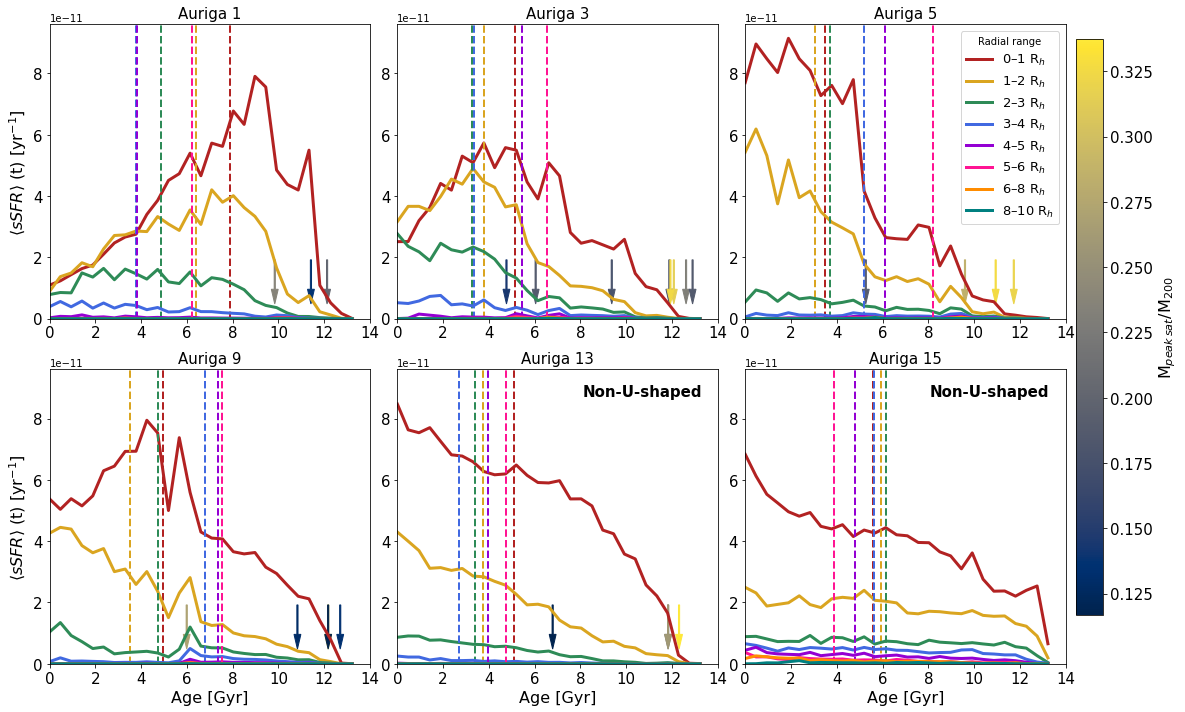}
    \caption{$⟨\rm sSFR⟩(t)$ of the in situ stellar populations at different radial ranges, as a function of their age. The arrows represent the merger events that satisfy that the total peak mass of the satellite galaxy was at least $\frac{1}{10} \, M_{200}$, considering the $M_{200}$ of the host galaxy at the time at which the satellite crossed its $R_{200}$ for the first time. They are colour-coded by the ratio between $M_{peak \, sat}$ and $M_{200}$. The vertical dashed lines represent the mean age at the different radii (from $0-1 \, R_h$ to $5-6 \, R_h$), and their colours match those of their corresponding radial ranges.}
    \label{fig:sSFRwmergers}
\end{figure*}

A possible mechanism that plays an important role in the redistribution of stellar material, and thus could be behind these U-shaped age profiles, is known as radial migration. This process allows stars to move through the disc plane to different galactocentric radii without significantly increasing their orbital eccentricities. 
One of the key drivers of radial migration is the interaction of stars with transient spiral arms. In this scenario, stars can exchange angular momentum with the spiral pattern, leading them to migrate inwards or outwards while remaining on nearly circular orbits \citep{Sellwood2002, Roskar2008a, Roskar2008b}. To explore if this mechanism plays a role in our galaxy sample, we computed the $m=2$ Fourier mode, which corresponds to a two-armed spiral structure \citep{Grand2016}. This analysis revealed that our galaxies have typically small amplitude of the A2 mode ($\lesssim 0.1$ at $z=0$), and that it has been generally very low within the last $\sim 8$ Gyr. This implies that this feature was not an important driver of stellar population radial mixing in the past either. We thus rule out the influence of spiral arms and radial migration in the rising of the U shape of the galaxies' radial age profiles. 

\subsection{The impact of merger events in the radial age profiles} \label{subsec:mergers}

An important external mechanism by which stars can be redistributed over time consists of interactions and mergers with satellite galaxies. If the galaxies with U-shaped age profiles underwent merger events at earlier times, these events could have redistributed the older stellar material, making it reach the outskirts of the galaxies. Thus, we analysed the merger events undergone by each galaxy involving satellite galaxies that reached a maximum mass ($M_{peak \, sat}$) of at least $\frac{1}{10} \, M_{200}$, where the $M_{200}$ of the main galaxy represents the total mass enclosed in a sphere with a mean density of 200 times the critical density of the Universe. In this analysis, we consider the $M_{200}$ at the time at which the satellite galaxy crossed the galaxy's $R_{200}$ for the first time. In Fig. \ref{fig:sSFRwmergers}, we show the temporal evolution of the time-averaged specific star formation rate, $⟨\rm sSFR⟩(t)$, of the in situ stellar populations at different radial ranges based on the $R_h$ at $z=0$ of each galaxy, for the subsample of 6 galaxies that we chose as examples. These profiles are reconstructed from the stellar age distribution at $z=0$ by computing the stellar mass formed within successive age bins, dividing by the bin width, and normalising by the total stellar mass of the galaxy at $z=0$. This same plot is shown for all galaxies in Fig. \ref{fig:sSFR} of Appendix \ref{appendixB}. The colour-coded vertical dashed lines represent the mean age of the stellar particles within radial bins up to $6 \, R_h$. On each panel, we highlight the significant merger experienced by each galaxy with colour-coded arrows, based on the ratio $M_{peak \, sat}/M_{200}$. Most galaxies (all but Auriga 15) have suffered important merger events, i.e. with merger ratios between $0.1$ and $\sim 1$ when selecting satellite galaxies that satisfy that its $M_{peak \, sat}$ is at least $\frac{1}{10} \, M_{200}$. We find that galaxies with U-shaped age profiles undergo merger events both at early and late times. However, most galaxies which do not show U-shaped age profiles, also accrete satellites as massive as some of the satellites accreted by the other galaxies with this feature in their profiles, besides undergoing merger events at similar times (see, for instance, Auriga 13 in Fig. \ref{fig:sSFRwmergers}). Hence, merger events alone are not enough to explain this behaviour in the age profiles, but instead this information must be combined with that of the $⟨\rm sSFR⟩(t)$ of the galaxies.

When a galaxy has already formed a considerable amount of stellar material at early times, this old in situ population can be redistributed outwards during early mergers. The subsequent inside-out star formation gives rise to the U shape found at $z=0$ in its radial age profile, where the inner regions present a negative age gradient until it reaches the location where there is no more recent star formation (see Fig. \ref{fig:sfrprof}), and thus the old stellar material that was ejected dominates. From that radius onward, the radial age profile presents an upturn that reflects the older ages found at those large radii. The galaxies with a U-shaped age profile, exemplified in Fig. \ref{fig:sSFRwmergers} with Auriga 1, 3, 5 and 9, had formed sufficient stellar material at early epochs (or by the merger time of the accreted satellites) in order to be redistributed during interactions with satellite galaxies. Additionally, the accreted stellar material that merger events bring is typically old (see Fig. \ref{fig:ageprof} and Fig. \ref{fig:AgeProfInsAcc}). When this material is deposited in the outer regions of the galaxy, it contributes to increasing the average stellar age at large galactocentric distances. This additional population of old accreted stars can therefore enhance the upward trend observed in the outer parts of the age profiles, reinforcing the U-shaped behaviour.

In the case of the galaxies without a U-shaped age profile, we note a clear difference in these galaxies' $⟨\rm sSFR⟩(t)$ with respect to those that have this feature. For instance, Auriga 13 and Auriga 15 in Fig. \ref{fig:sSFRwmergers} show a continuous $⟨\rm sSFR⟩(t)$ through the years that increases monotonically until $z=0$ (i.e., ages $\sim 0$ Gyr). This means that these galaxies are continuously forming stellar material. Moreover, in some cases these galaxies also sustain star formation at larger radii than those with U-shaped profiles, as seen for example in Auriga 15. The presence of ongoing star formation across all radii, combined with their steadily increasing sSFR, results in either flat or decreasing radial age profiles due to the presence of young stellar populations throughout the galaxy. 
We also note that Auriga 15 does not suffer any merger as massive as those galaxies with a U-shaped radial age profile that can significantly impact its distribution of stellar material.

We therefore conclude that the cessation of star formation beyond certain radius and the interactions with satellite galaxies are the most dominant mechanisms responsible for the U shape in the central parts of the radial age profiles, while radial stellar migration driven, for example, by spiral arms has no impact on the matter. The merger events redistribute stars outwards, reversing the negative age gradient within the inner disc and contributing to a positive gradient in its outer regions. This trend continues throughout the outskirts of the galaxies (i.e. beyond $4 \, R_h$), where they are dominated by old stellar populations that also include accreted stellar material, thereby preserving the overall age increase at large radii. 

Regarding the effect on the galaxies' metallicity, this redistribution of stellar material also has an impact in their radial [Fe/H] profiles given that the outskirts become dominated by older, more metal-poor stellar populations compared to the younger, more enriched centres, and generates the negative metallicity gradients shown in Sect. \ref{sec:metanalysis}. Moreover, as in situ populations of different ages are redistributed to large radii, the resulting mixture of stellar ages enhances the flattening of the metallicity profiles in the outer regions of the galaxies that is already influenced by the different metallicities of the accreted satellites that deposited their material in the outskirts.


\section{Discussion} \label{sec:discussion}

In this section, we place our results in an observational context, and we also compare them to those found in the MW-mass regime.

\subsection{Metallicity of low-mass galaxies} \label{subsec:discussion_met}

A wide variety of observed metallicity gradients have been reported when considering the low-mass regime of galaxies \citep[e.g.][]{Leaman2013, Taibi2018, Fu2024a, CanoDiaz2025, Sato2025}, even including galaxies with no metallicity gradients at all \citep[e.g.][]{Aparicio2001, Taibi2024}. The metallicity gradients for low-mass galaxies obtained in this work fall well within the observed values for isolated dwarfs \citep[e.g.][]{Taibi2020, Taibi2022, Fu2024b}. We do not find any significant correlation between the metallicity gradients of the galaxies in our sample and either their total stellar mass or their accreted stellar mass (see Fig. \ref{fig:MetGrad_vs_Mstel}). This is in agreement with the findings of \cite{Taibi2022}, where no correlation is found between the metallicity gradient and the stellar mass of the isolated dwarf galaxies of their sample. This implies that the link between the evolution of the galaxies and their resulting metallicity gradients may not be so straightforward to measure in the low-mass regime. Similarly, when considering only the metallicity gradients of the stellar halos' both total and accreted components, we also find no clear correlation with their respective stellar halo masses. However, \cite{Taibi2022} find a mild correlation between these gradients and the luminosities of the galaxies. We therefore checked whether this correlation is also present in our simulated galaxies. The luminosities of our galaxies range from $5.3 \times 10^6 \, L_\odot$ to $1.9 \times 10^8 \, L_\odot$ (see Table \ref{tab:MetGrad_L}), which are consistent with those observationally measured by \cite{Taibi2022} that range from $\sim 10^5 \, L_\odot$ to $\sim 10^8 \, L_\odot$. We found no correlation between the metallicity gradients of the galaxies measured in dex/R$_h$ and their luminosities, but we did find a mild correlation between the metallicity gradients computed in physical units (dex/kpc) for the whole galaxy and the main body of the galaxies, and their luminosities (see Appendix \ref{app}). We show this trend in Fig. \ref{fig:MetGrad_kpc_TaibiComp}, as well as the comparison with the observational data taken from \cite{Taibi2022} for isolated (filled squares) and satellite (empty squares) LG dwarf galaxies. We note that the metallicity gradients measured in dex/kpc and the luminosity values of our simulated galaxies are in agreement with those observationally measured by \cite{Taibi2022} for isolated dwarf galaxies.

\begin{figure}[!ht]
   \centering
   \includegraphics[width=\columnwidth]{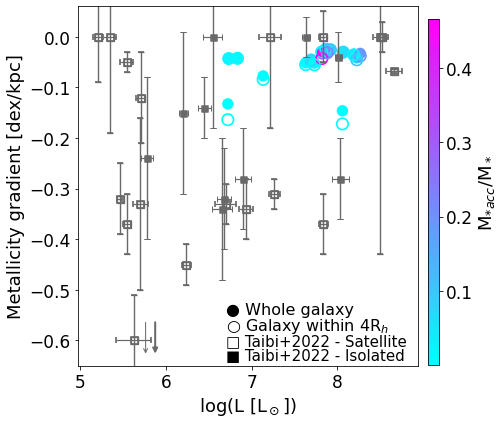}
    \caption{Metallicity gradient in physical units (dex/kpc) of the galaxies of our sample computed for the main body of the galaxies (empty circles) and for the whole galaxy (filled circles), as a function of their luminosities and colour-coded by the galaxies' total accreted mass fraction. Observational data taken from \cite{Taibi2022} for isolated (filled squares) and satellite (empty squares) LG dwarf galaxies are also shown. The filled (empty) arrow represents the isolated (satellite) galaxy Phoenix (Leo II), which has a metallicity gradient of $\sim -1.3$ ($\sim -1.4$) [dex/kpc]. We note that the metallicity gradient and the luminosity values of our simulated isolated galaxies are in good agreement with those observationally measured for isolated dwarf galaxies.}
    \label{fig:MetGrad_kpc_TaibiComp}
\end{figure}

As previously shown in \hyperref[T25ref]{T25}, our results demonstrate a clear correlation between the mass and the metallicity of the accreted stellar halos in low-mass galaxies, in which more massive accreted stellar halos are also more metal rich (Fig. \ref{fig:HaloFeH}). This extends the stellar halo mass–metallicity relation found for MW-mass galaxies \citep{Harmsen2017, Monachesi2019} to the low-mass regime. In this work, we analysed the spread in metallicity that accreted stellar halos with a similar mass have, and we found that it can be attributed to the infall time of the most dominant satellite they accreted. This is due to the fact that a later infall time at similar satellite mass implies that the satellites had more time to enrich their stellar content before disruption. Therefore, the accreted component of the stellar halos with a later infall time of their dominant satellites will be more metal-rich than those with the same amount of accreted stellar mass that accreted theirs earlier. This interpretation aligns with observational results that point to late massive accretion events being responsible for the relatively metal-rich halos of MW-mass galaxies like M31 \citep[e.g.,][]{Gilbert2014, DSouza2018}. Our results therefore suggest that similar mechanisms operate at lower masses, and highlight that the metallicities of the accreted halos of dwarf galaxies do retain information of the timing of their most significant accretion events (see also Gonzalez-Jara et al., in prep.).

\subsection{U-shaped age profiles in dwarf galaxies} \label{subsec:discussion_age}

In this work we have found that most low-mass galaxies of our sample present a U-shaped age profile (Fig. \ref{fig:ageprof}), suggesting intricate star formation and evolution. In contrast, MW-mass galaxies typically display negative age gradients, consistent with inside-out star formation \citep[e.g.][]{Sanchez-Blazquez2014, GonzalezDelgado2015}, although U shapes have been observed in some cases \citep[e.g.][]{Yoachim2012}. Some theoretical studies have predicted that this U-shaped behaviour is possible in such massive systems. For example, \cite{Roskar2008b} simulate a disc galaxy with a Navarro-Frenk-White DM halo \citep{Navarro1997} of $10^{12} \, M_\odot$, without a bar or strong spiral arms at $z=0$, and find a clear U-shaped age profile, attributed to transient spiral arms reshuffling stellar material. Similarly, \cite{Sanchez-Blazquez2009} use N-body simulations of a disc galaxy of dynamical mass $7.6 \times 10^{11} \, M_\odot$, and also find a U-shaped age profile. They argue that, in addition to radial migration, the U shape arises from a break in star formation density: beyond this radius, star formation declines sharply, resulting in older disc outskirts, which is consistent with observational evidence that star formation becomes inefficient below a critical gas surface density as described by the Kennicutt–Schmidt relation in both MW–mass and dwarf galaxies \citep[e.g.][]{Leroy2008, Kennicutt2012}. They also claim that accretion events have little impact on this gradient. Furthermore, \cite{VarelaLavin2022} analysed 1012 disc galaxies from the EAGLE simulations ($10^{10} \leq M_* \leq 10^{11.5} \, M_\odot$) and reported that U-shaped age profiles are frequent in their sample.

In low-mass galaxies, U-shaped age profiles appear more often in observations. For example, \cite{Williams2009} used data from the Hubble Space Telescope (HST)/Advanced Camera for Surveys (ACS) to study the disc of M33 and found an age inversion: 
in the inner disc, the stellar age decreases with increasing distance from the centre, contrasting with the outer disc where older stars are found. Furthermore, \cite{Pessa2023} observed a U-shaped radial age profile in two of the $19$ star-forming galaxies analysed in their work, NGC 1087 and NGC 1385, which are also within the stellar mass range analysed in our work ($\log(M_*/M_\odot) = 9.9$ and $\log(M_*/M_\odot) = 10$, respectively). More recently, a U-shaped age profile has also been reported in the Large Magellanic Cloud (LMC). \cite{Cohen2024b} argued that the inversion they found in its radial age profile cannot be explained solely by the interaction between the LMC and the Small Magellanic Cloud (SMC) due to the fact that the radial location of this inversion remains constant over a wide range of look-back times, suggesting that the mechanisms influencing star formation in the LMC potentially predate recent interactions with the SMC. 

The results obtained in this work regarding the radial mean age profiles of low-mass galaxies can help shed light on these aforementioned observations of such systems. First of all, according to the predictions made by our simulations, it is likely that the U-shaped age profiles observed in the central parts of M33, NGC 1087, NGC 1385, and the LMC is caused by in situ material redistributed to the outer regions. However, regarding M33, our results do not seem to agree with those of \cite{Mostoghiu2018}, where the authors analysed an analogue of M33 from the CLUES simulation and found that the main driver of the mean age profile's U shape is the accreted component of said galaxy, brought in by minor mergers. While we show here that the accreted material is indeed old and contributes to the older ages in the outer regions, this accreted material is not responsible for the U shape in the age profiles, which is generated by in situ material (see Fig. \ref{fig:AgeProfInsAcc}). They also found that the stellar migration of old in situ stellar populations from the central regions towards the outskirts of M33 help shape the radial mean age profile, although to a much lesser extent compared to the contribution of the accreted stellar material. 

Secondly, based on our results, we suggest that the U shape in the radial age profiles of these observed galaxies arises from the absence of recent star formation beyond the radius where the stellar populations are youngest, together with the redistribution of their old stellar populations due to mergers and interactions. In the case of the LMC, \cite{Cohen2024b} demonstrated that the inversion in the galaxy's age profile approximately matches the drop-off in the HI column density, which agrees with the results of \cite{Meschin2014} that argue that the HI disc of the LMC able to form stars has indeed shrunk at these distances, thus likely limiting star formation in the outer regions. Similarly, recent work by \cite{Corbelli2025} on M33 shows that while the galaxy continues to form stars across its entire star-forming disc, the star formation rate declines significantly with radius, particularly beyond the inner disc, due to variations in gas density, stability, and disc dynamics. Together, these observational findings support the plausibility of our proposed explanation for the formation of U-shaped age profiles in low-mass galaxies. 
We bear in mind that our numerical predictions are based on isolated galaxies at $z=0$, but similar processes can also operate in satellite low-mass systems. On one hand, denser environments foster frequent interactions and tidal perturbations that can redistribute stellar material. On the other hand, the intrinsic evolution of each galaxy, governed by its star formation history, will happen in galaxies regardless of their environment. Thus, the mechanisms we propose remain relevant for shaping radial age profiles even in dense environments.

Another galaxy that could have a U-shaped age profile is NGC 7793, a relatively isolated low-mass \citep[$M_* \approx 3.32 \times 10^9 M_\odot$,][]{Dale2009} spiral galaxy. \cite{Sacchi2019} found an inside-out growth in the inner region of its disc. However, it is not possible to confirm from their data whether this inside-out trend persists across the entire disc, particularly in the outermost regions, where older stellar populations than those found closer to the centre were reported. This inner inside-out growth is also confirmed by \cite{Kang2023}, and their reported age profile appears to exhibit either a flattening or a reversal at large radii along the disc. Unfortunately, their analysis, similarly to that of \cite{Sacchi2019}, does not cover farther distances. Therefore, more observational data covering large distances from the centre of NGC 7793 could help determine whether this galaxy has a U-shaped radial age profile as well.

Our findings can also be compared with other numerical predictions. \cite{Graus2019} used the Feedback In Realistic Environments (FIRE) to study 26 dwarf galaxies with stellar masses ranging from $\sim 10^5 \, M_\odot$ to $\sim 10^9 \, M_\odot$, and found a prevalence of outside-in growth, with the galaxies showing younger stellar populations in their centres than in their outskirts. They attribute these positive age gradients to a combination of mergers and stellar heating, noting that while mergers alone are not particularly effective at shaping the gradients, they can still contribute to some redistribution of stars, which is a result consistent with our own findings. Regarding the redistribution of stars, they explain this with heating by stellar feedback, which strongly affects the galaxies in the FIRE simulations \citep[e.g.][]{Onorbe2015, Fitts2017}. Stellar feedback generates potential fluctuations that dynamically heat stars and drive molecular outflows which subsequently form new (and thus younger) stars \citep{ElBadry2016}. These feedback episodes promote the outward migration of older stellar populations to larger radii (see also \citealt{Riggs2024}). Thus, even though \cite{Graus2019} primarily attribute the reshuffling of old stellar material to stellar feedback-driven migration rather than the mechanisms we propose with the Auriga simulations, both works highlight the important role of mergers in contributing to stellar redistribution.

\subsection{Stellar halos age and metallicity} \label{subsec:discussion_agemet}

While the correlation between age and metallicity in stellar halos has been explored in detail for the MW \citep[e.g.][]{Jofre2011, Schuster2012}, it remains uninvestigated in the low-mass regime because these substructures have just recently started to be detected around low-mass systems. Observational work on dwarf galaxies has primarily focused on characterising metallicity gradients within their main bodies and overall chemical abundance patterns, as discussed in Sect. \ref{subsec:discussion_met}, often without simultaneously resolving detailed age distributions in their stellar halos. This is due to the faintness of these structures that makes them difficult to observe at large distances.

Thus, our numerical analysis offers insights into the unexplored age–metallicity relationship in the stellar halos of dwarf galaxies, revealing a clear correlation between these properties in their accreted component. Confirming such a trend observationally remains challenging at present, which underscores the predictive value of our results. These suggest that, once observations can robustly recover age distributions in dwarf galaxy stellar halos, a trend should emerge in which more metal-rich accreted stellar halos are systematically younger. In the meantime, our results provide a framework for inferring a likely age range for the stellar halo of low-mass galaxies based on its measured metallicity or mass, offering a practical tool to interpret existing observational data and to guide future targeted studies. However, observations provide data of the total stellar halo, and as seen in \hyperref[T25ref]{T25}, the inner regions of these structures are dominated by in situ stellar material. Therefore, developing robust methods to disentangle the accreted and in situ components in observed stellar halos is essential to fully exploit these predictions. One way of doing so could be with a cut in [Fe/H] values, considering that those lower than $-1$ will correspond to the accreted component, given that the total stellar halos median [Fe/H] values range from $\sim -1$ to $\sim -0.4$ (see Fig. \ref{fig:HaloFeH}).

\section{Summary and conclusions} \label{sec:conclusions}

In this work we have used $17$ low-mass galaxies ($\sim 3 \times 10^8 \, M_\odot \leq M_* \lesssim 2 \times 10^{10} \, M_\odot$) available in the Auriga project \citep{Grand2024} to study the ages and metallicities of such systems, spanning from their centres to their stellar halos. These latter structures are defined as the stellar material of the galaxies located outside an ellipsoid with semi-major axes equal to $4 \, R_h$ and up to $10 \, R_h$ in all directions, where $R_h$ is the half-light radius of the galaxy. The main conclusions can be listed as follows:

   \begin{enumerate}
      \item The galaxies analysed in this work have negative [Fe/H] gradients, ranging from $-0.18$ to $-0.07$ in units of dex/$R_h$, indicating that their outskirts are more metal-poor than their central regions. These gradients are defined as the slopes in the variations of the median [Fe/H] values as a function of the radius normalised to $R_h$. The centres of less massive dwarfs ($M_* < 6.3 \times 10^9 \, M_\odot$) are in general more metal-poor than those of more massive dwarfs.
      \item We find no correlation between the metallicity gradients in dex/$R_h$ of the galaxies and their stellar mass, their accreted stellar mass, their accreted mass fraction, and the number of significant progenitors they have. This implies that the link between the evolution of the galaxies and their resulting metallicity gradients is not so straightforward to measure in the low-mass regime. However, we find a mild correlation between the metallicity gradients and the luminosities of the galaxies when the gradients are computed in physical units (i.e. dex/kpc).
      \item More massive accreted stellar halos ($M_{acc \, halo} \geq 10^8 \, M_\odot$) are also more metal-rich (median [Fe/H] values $\geq -1.22$), following the mass-metallicity relation of stellar halos also found in MW-mass galaxies. We find a correlation between the metallicity of the accreted stellar halos and the infall time of their most dominant satellite, which translates into a dispersion in the mass-metallicity relation: at a given accreted stellar halo mass, the dwarf galaxies that accreted their most dominant progenitor at later times have a more metal-rich accreted stellar halo. This is due to the fact that satellites that were accreted at later times had more time to chemically evolve and enrich their stellar populations before being disrupted.
      \item There is a strong correlation between the age and metallicity of the accreted stellar halos of the analysed low-mass galaxies: more metal-rich accreted stellar halos are younger, while more metal-poor ones are older.
      \item The mean ages of the main body of the galaxies (i.e. within $4 \, R_h$) are generally lower than the mean ages of their total stellar halos (see Table \ref{tab:properties}).
      \item More massive stellar halos are older than the less massive ones, indicating that less massive galaxies keep on forming in situ stellar material until later times, which can later end up comprising their stellar halos. In contrast, when considering only the accreted component, more massive accreted stellar halos have younger stars than less massive ones, given that the less massive galaxies they surround accreted their stellar material earlier, leaving them with older accreted material in their stellar halos.
      \item The majority of the galaxies (12 out of 17) present a prominent U shape in their radial age profiles. This U shape is mainly driven by the in situ stellar population and located within $4 \, R_h$.
      \item In those galaxies with a U-shaped radial age profile, the radial location of the minimum age before the upturn matches that of the peak of the star formation of young stellar material. In other words, in cases where we found a U-shaped radial age profile, the age of the stellar material starts to radially increase beyond the radius where there is no recent star formation activity.
      \item Galaxies with and without a U shape in their radial age profile undergo significant merger events (i.e. with $M_{peak}/M_{200}$ ratios between $0.1$ and $1$ when considering satellite galaxies that satisfy $M_{peak} \geq \frac{1}{10} \, M_{200}$) at early times and late times. Hence, these alone are not enough to explain the redistribution of the old stellar material so as to generate the U-shaped radial age profiles. However, when this information is combined with the sSFR, we note that galaxies with a U shape had formed a considerable amount of stellar material at early times that was already there to be kicked out when the mergers occurred. Furthermore, the galaxies without a U-shaped age profile present a continuous sSFR that increases monotonically up to $z=0$, and they keep on forming stellar material at large distances from their centres, which contributes to the flattening of their profiles. 
   \end{enumerate}

These results show us the wide variety in ages and metallicities when analysing low-mass galaxies and their stellar halos, reflecting the complex and non-uniform evolutionary pathways these systems can follow. The trends we identify point to significant variations in their accretion histories, with differences in the timing, mass, and chemical composition of accreted satellites leaving distinct imprints on their present-day properties. Although these imprints on stellar halos are difficult to confirm with current observational data, future facilities such as the LSST and the Nancy Grace Roman Space Telescope will enable the detection and characterisation of these extended components.

\begin{acknowledgements}
      We thank the anonymous referee for their suggestions, which helped improve this manuscript. EAT would like to thank Azadeh Fattahi and Victor Debattista for insightful discussions. EAT acknowledges financial support from ANID ``Beca de Doctorado Nacional” 21220806 and from the Centro de Excelencia en Astrofísica y Tecnologías Afines (CATA). EAT, AM, and FAG acknowledge support from the ANID BASAL project FB210003. AM and FAG acknowledge support from the ANID FONDECYT Regular grant 1251882 and 1251493, respectively, and funding from the HORIZON-MSCA-2021-SE-01 Research and Innovation Programme under the Marie Sklodowska-Curie grant agreement number 101086388. RJJG acknowledges support from an STFC Ernest Rutherford Fellowship (ST/W003643/1). FvdV is supported by a Royal Society University Research Fellowship (URF R 241005). RB is supported by the SNSF through the Ambizione Grant PZ00P2-223532. We have used simulations from the Auriga Project public data release \citep{Grand2024} available at https://wwwmpa.mpa-garching.mpg.de/auriga/data.html.
\end{acknowledgements}

\bibliography{Bibliography}

\begin{appendix}

\section{Comparison with a cylindrical geometry} \label{appendix0}

\begin{figure}[!ht]
   \centering
   \includegraphics[width=\columnwidth]{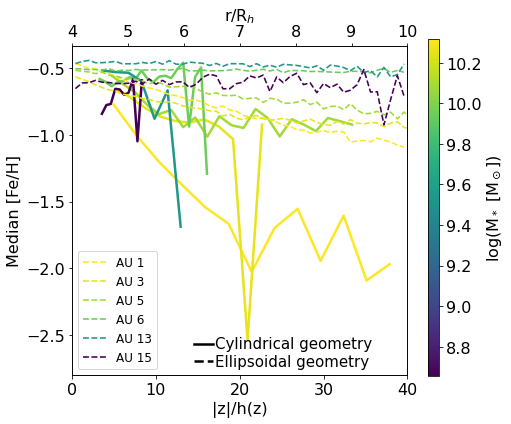}
    \caption{Metallicity profiles of the stellar halo of a subsample of galaxies. The solid profiles were computed using a cylindrical geometry, while the dashed ones were computed using the original ellipsoidal geometry explained in Sec. \ref{subsec:metallicity}. The cylindrical profiles are normalised by the scale height of the galaxies, while the ellipsoidal profiles are normalised by their $R_h$. All profiles are colour-coded by the galaxies' stellar mass.} 
    \label{fig:GeometricalComparison}
\end{figure}

We re-measured the median [Fe/H] profiles for all galaxies using a cylindrical geometry, measuring vertical metallicity profiles, mimicking minor axis profiles. The stellar halo region for the minor axis profile was defined adopting a maximum cylindrical radius of $1 \, R_h$ and excluding particles within the semi-minor axis c to minimise disc contamination (i.e. the minimum considered |z| value to define the cylinder was that of c, listed in Table \ref{tab:properties}). The maximum |z| value considered for this geometry was $10 \, R_h$, to be consistent with our stellar halo definition. Fig. \ref{fig:GeometricalComparison} shows these new profiles computed with a cylindrical geometry for the stellar halo (solid lines) of a subsample of galaxies, as well as those obtained with the original ellipsoidal geometry (dashed lines) using $4 \, R_h$ and $10 \, R_h$ as the inner and outer limits of this structure, respectively. We find that the resulting vertical metallicity profiles of the stellar halo are overall not always flat, decreasing with galactocentric distance in most cases. This indicates that our results are affected by the adopted geometry. 
This cylindrical approach is particularly advantageous for comparisons with observational data, as studies of the stellar halo often rely on measurements taken along the semi-minor axis.

\section{Metallicity gradients in physical units} \label{app}

\begin{figure*}[!ht]
   \centering
   \sidecaption
   \includegraphics[width=12cm]{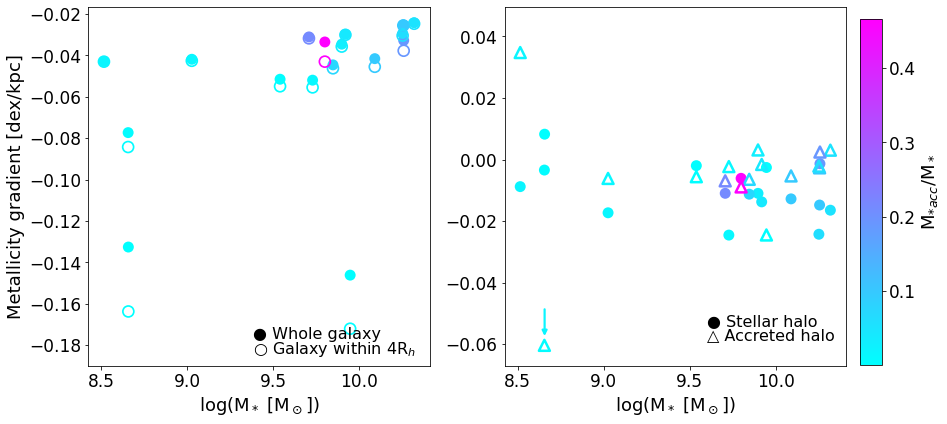}
    \caption{Left panel: Metallicity gradient in physical units (dex/kpc) of the galaxies of our sample computed within $4 \, R_h$ (empty circles) and $10 \, R_h$ (filled circles), as a function of their stellar mass. Right panel: Metallicity gradients in physical units (dex/kpc) of the total stellar halo and the accreted stellar halo as a function of the galaxies' stellar mass, represented with filled circles and empty triangles, respectively. The arrow represents the value corresponding to the accreted stellar halo of Auriga 15, which is $\sim -0.26$ [dex/kpc]. Both panels are colour-coded by the galaxies accreted mass fraction.} 
    \label{fig:MetGrad_dexkpc}
\end{figure*}

We also computed the metallicity gradients for the whole galaxy, the main body of the galaxies, and the total and accreted stellar halos of the galaxies in units of dex/kpc instead of dex/$R_h$. The resulting values are shown as a function of the galaxies' stellar mass in Fig. \ref{fig:MetGrad_dexkpc}, colour-coded by their accreted mass fraction. We note a moderate correlation between these quantities when considering the whole galaxy and its main body, that was further quantified using a Spearman correlation coefficient, which resulted in $\rho = 0.64$ for the whole galaxy and $\rho = 0.56$ for its main body.

Additionally, for the main body of the galaxies (and also for the whole galaxy), we find that the values of these metallicity gradients in dex/kpc are consistent with the previous observational findings of \cite{Taibi2022}, which gradients span a range of values from -1.3 dex/kpc to ~0 dex/kpc, with the majority of the values of their sample found between -0.5 dex/kpc and ~0 dex/kpc.

\begin{figure*}[!ht]
   \centering
   \sidecaption
   \includegraphics[width=12cm]{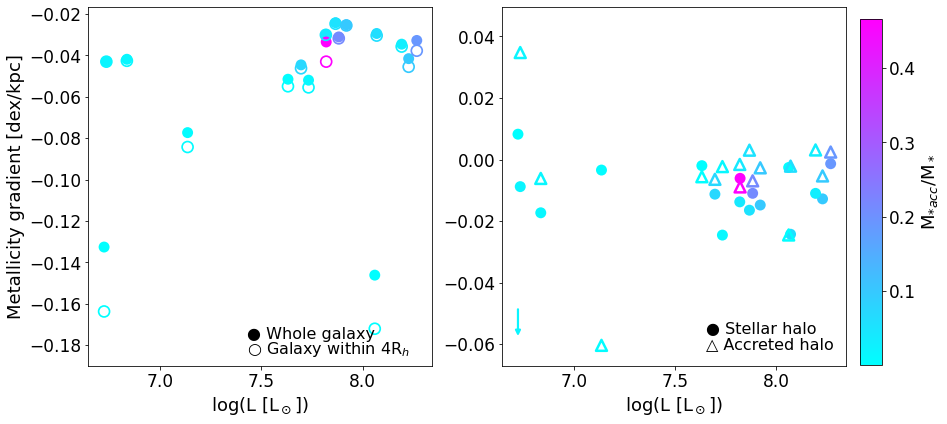}
    \caption{Left panel: Metallicity gradient in physical units (dex/kpc) of the galaxies of our sample computed for the main body of the galaxies (within $4 \, R_h$, empty circles) and for the whole galaxy (within $10 \, R_h$, filled circles), as a function of their luminosities. Right panel: Metallicity gradients in physical units (dex/kpc) of the total stellar halo and the accreted stellar halo as a function of the galaxies' luminosity, represented with filled circles and empty triangles, respectively. The arrow represents the value corresponding to the accreted stellar halo of Auriga 15, which is $\sim -0.26$ [dex/kpc]. Both panels are colour-coded by the galaxies' accreted mass fraction.} 
    \label{fig:MetGradkpc_L}
\end{figure*}

Given that \cite{Taibi2022} reported a mild correlation between the galaxies metallicity gradients and their luminosities, we tested if this correlation is also found in the Auriga simulations. We find no correlation between the gradients and the luminosities when considering the metallicity gradients computed in units of dex/R$_h$. However, we find a mild correlation when considering the metallicity gradients of the whole galaxy and of its main body computed in physical units. We show these quantities as a function of the galaxies' luminosities in Fig. \ref{fig:MetGradkpc_L}. We note this moderate correlation only when considering the whole galaxy (left panel), but none when considering the stellar halo or the accreted stellar halo. We further corroborated this mild correlation with a Spearman correlation coefficient, which indeed confirmed its existence.

We present in Table \ref{tab:MetGrad_L} the computed values for the luminosity of our galaxies, and the metallicity gradients in physical units for the whole galaxies and for the main bodies of the galaxies.

\begin{table}[!ht]
    \centering
    \caption{Metallicity gradients in physical units and luminosities of the galaxies analysed in this work.}
    \label{tab:MetGrad_L}
    \begin{tabular}{lccr}
    Auriga &    Luminosity &  $\nabla$[Fe/H]$_{<\, 10 \, R_h}$ &  $\nabla$[Fe/H]$_{<\, 4 \, R_h}$ \\
     & [$L_\odot$] & [dex/kpc] & [dex/kpc] \\
    \hline
     1 &  $7.35 \times 10^7$ &        -0.02 &           -0.02 \\
     2 &  $1.85 \times 10^8$ &        -0.03 &           -0.04 \\
     3 &  $8.31 \times 10^7$ &        -0.03 &           -0.03 \\
     4 &  $1.17 \times 10^8$ &        -0.03 &           -0.03 \\
     5 &  $1.69 \times 10^8$ &        -0.04 &           -0.05 \\
     6 &  $1.15 \times 10^8$ &        -0.15 &           -0.17 \\
     7 &  $6.58 \times 10^7$ &        -0.03 &           -0.03 \\
     8 &  $1.56 \times 10^8$ &        -0.03 &           -0.04 \\
     9 &  $4.96 \times 10^7$ &        -0.04 &           -0.05 \\
    10 &  $6.61 \times 10^7$ &        -0.03 &           -0.04 \\
    11 &  $5.40 \times 10^7$ &        -0.05 &           -0.06 \\
    12 &  $7.63 \times 10^7$ &        -0.03 &           -0.03 \\
    13 &  $4.28 \times 10^7$ &        -0.05 &           -0.05 \\
    14 &  $6.84 \times 10^6$ &        -0.04 &           -0.04 \\
    15 &  $5.28 \times 10^6$ &        -0.13 &           -0.16 \\
    16 &  $1.37 \times 10^7$ &        -0.08 &           -0.08 \\
    17 &  $5.42 \times 10^6$ &        -0.04 &           -0.04 \\
    \hline
    \end{tabular}
\end{table}

\section{Mean age profiles} \label{appendixA}

\begin{figure*}[!ht]
   \centering
   \includegraphics[width=2\columnwidth]{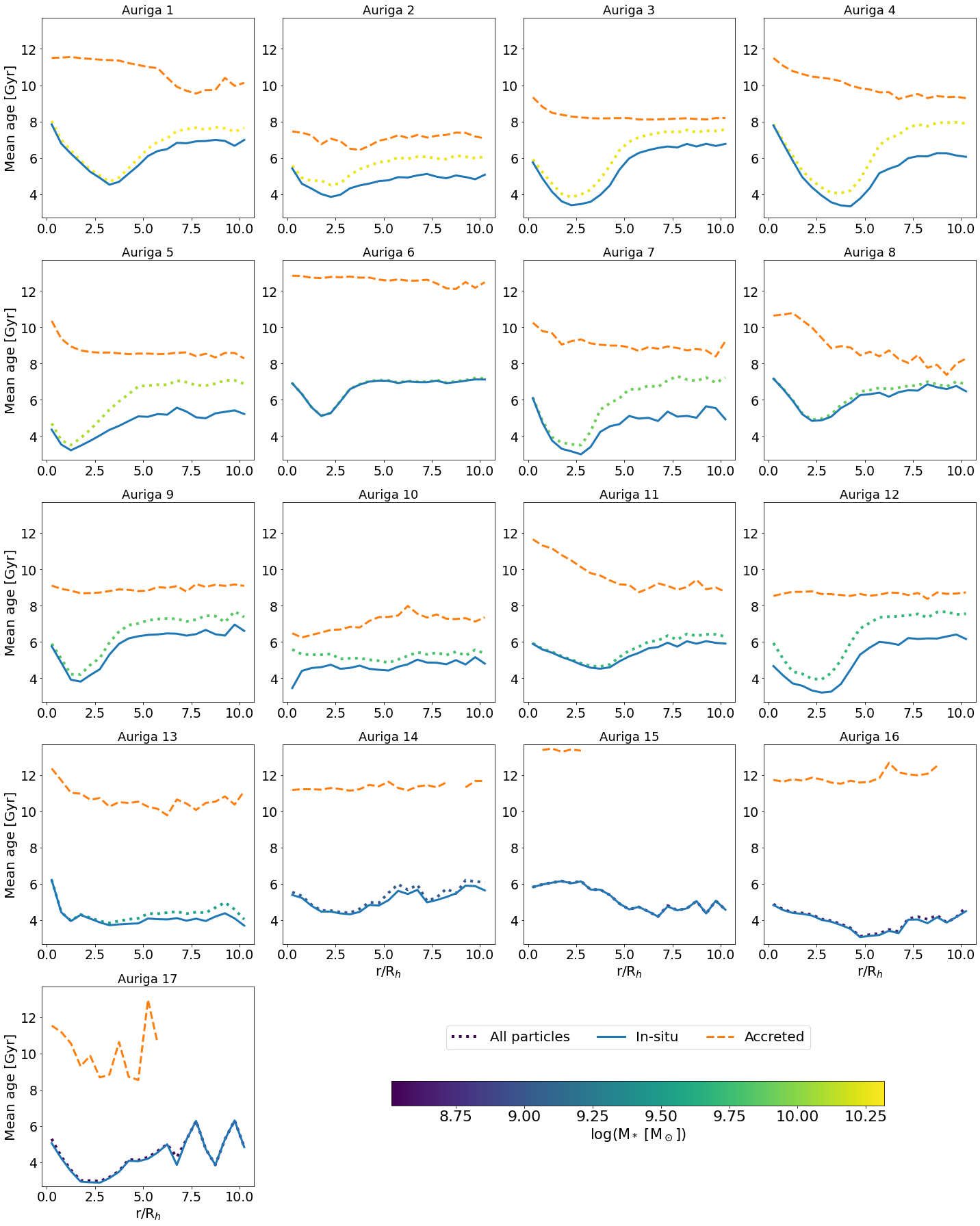}
    \caption{In situ (solid blue line), accreted (dashed orange line) and total (dotted line) mean age profiles of all galaxies in our sample. The total mean age profiles are coloured according to the galaxies' stellar mass. We note a prominent U shape in the profile of 12 galaxies, driven by the in situ stellar material.} 
    \label{fig:AgeProfInsAcc}
\end{figure*}

In Fig. \ref{fig:AgeProfInsAcc}, we show the in situ (blue solid line), accreted (orange dashed line) and total (dotted line) mean age profiles for all galaxies of our sample. We colour-coded the total mean age profiles according to the galaxies total stellar mass. As discussed in Sect. \ref{subsec:ageprof}, we find that 12 out of 17 galaxies present a U-shaped mean age profile, which can be attributed to the behaviour of the in situ stellar component. 

\section{sSFR} \label{appendixB}

\begin{figure*}[!ht]
   \centering
   \includegraphics[width=1.9\columnwidth]{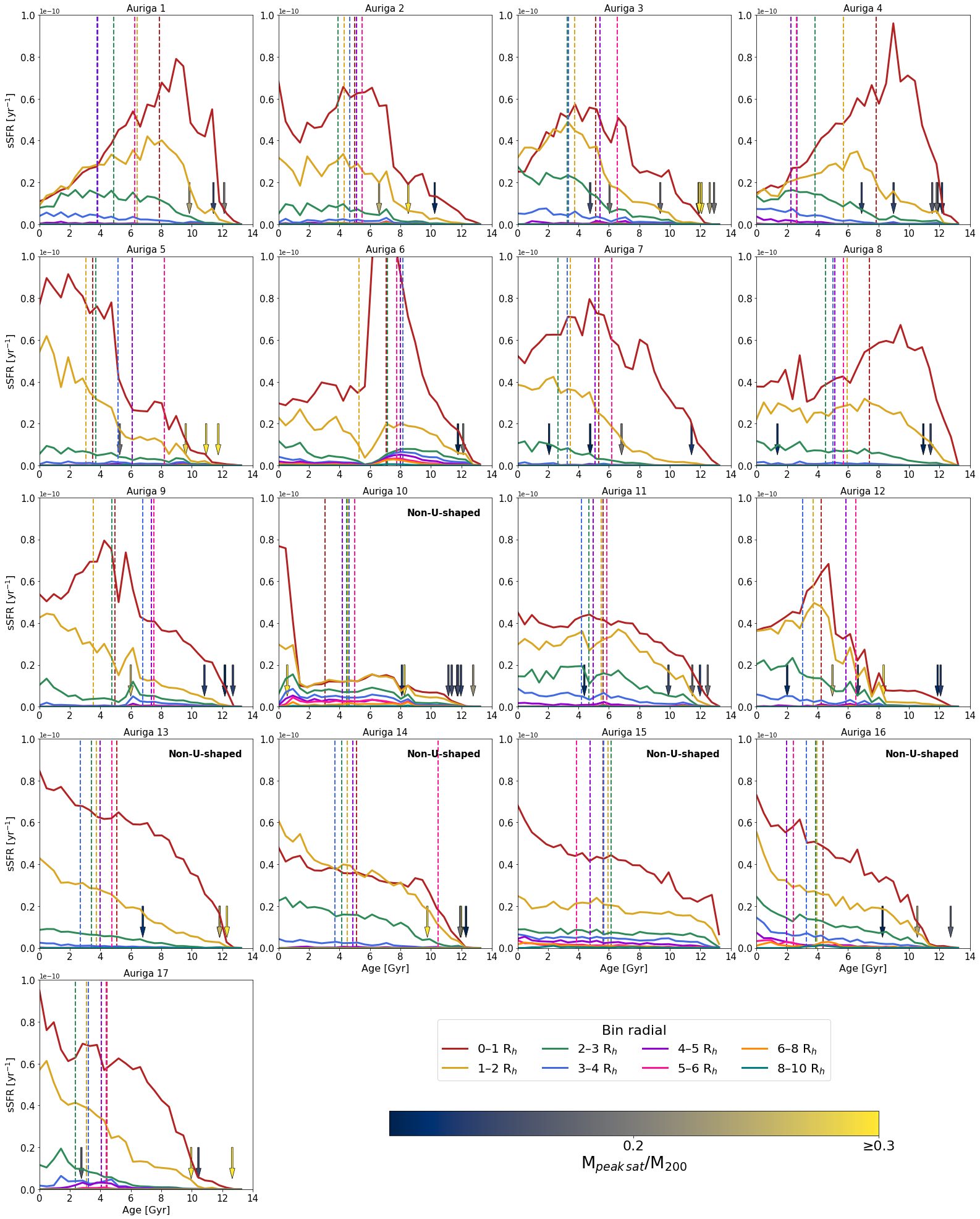}
    \caption{$⟨\rm sSFR⟩(t)$ of the in situ stellar populations at different radial ranges for all the galaxies of our sample, as a function of the age of the particles. The arrows represent the merger events that satisfy that the total peak mass of the satellite galaxy was at least $\frac{1}{10} \, M_{200}$, considering the $M_{200}$ of the host galaxy at the time at which the satellite crossed its $R_{200}$ for the first time. They are colour-coded by the ratio between the $M_{peak \, sat}$ and the $M_{200}$. The vertical dashed lines represent the mean age at the different innermost radii (from $0-1 \, R_h$ to $5-6 \, R_h$), and their colours match those of their corresponding sSFR.} 
    \label{fig:sSFR}
\end{figure*}

In Fig. \ref{fig:sSFR} we show the $⟨\rm sSFR⟩(t)$ for the in situ stellar particles' of the galaxies, as a function of their ages. We divide the stellar populations in different radial ranges, covering the whole galaxy from the centre to $10 \, R_h$. The vertical dashed lines represent the mean age at the different innermost radii, and their colours match those of the sSFR in the same radial bin. Merger events that satisfy that the total peak mass of the satellite galaxy was at least $\frac{1}{10} \, M_{200}$ are symbolised with arrows, colour-coded by the ratio between the $M_{peak \, sat}$ and the $M_{200}$. The merger ratios that are higher than $0.3$ are equally coloured. The considered $M_{200}$ of the host galaxy is the one it had at the time at which the satellite crossed its $R_{200}$ for the first time.

The case of the most recent merger suffered by Auriga 10 is discussed in \hyperref[T25ref]{T25} and we refer the reader to said work for details. Briefly, the galaxy accreted a satellite comparable in mass to its own at very late times, which is why we obtain that $M_{peak \, sat}/M_{200} = 1.09$. 

\end{appendix}

\end{document}